\documentclass[pdflatex,sn-mathphys-num]{sn-jnl}

\usepackage{graphicx}%
\usepackage{multirow}%
\usepackage{amsmath,amssymb,amsfonts}%
\usepackage{amsthm}%
\usepackage{mathtools}
\usepackage{mathrsfs}%
\usepackage[title]{appendix}%
\usepackage{xcolor}%
\usepackage{textcomp}%
\usepackage{manyfoot}%
\usepackage{booktabs}%
\usepackage{algorithm}%
\usepackage{algorithmic}%
\usepackage{listings}%
\usepackage{pgfplots}
\pgfplotsset{compat=1.18}

\usepackage{float}

\usepackage{letltxmacro}
\usepackage{xspace}
\usepackage{xcolor}
\usepackage{hyperref}
\hypersetup{hypertexnames=false}
\usepackage[colorinlistoftodos]{todonotes}
\usepackage[T1]{fontenc}

\theoremstyle{thmstyleone}%
\newtheorem{theorem}{Theorem}%

\newtheorem{definition}[theorem]{Definition}
\newtheorem{proposition}[theorem]{Proposition}

\newtheorem{lemma}[theorem]{Lemma}
\newtheorem{fact}[theorem]{Observation}

\newcommand{\GAP}{\text{--}}
\newcommand{\supp}{\mathop{\mathrm{supp}}}

\begin{document}

\title[Edge-wise Graph Alignments]{Edge-wise Graph Alignments}

\author*[1]{\fnm{Akbar} \sur{Davoodi}}\email{a.davoodi@heu.cas.cz}

\author[2]{\fnm{Christoph} \sur{Flamm}}\email{xtof@tbi.univie.ac.at}

\author*[3,4,5]{\fnm{Daniel} \sur{Merkle}}\email{daniel.merkle@uni-bielefeld.de}

\author[6,7,8,9,10]{\fnm{Peter F.} \sur{Stadler}}\email{studla@bioinf.uni-leipzig.de}

\affil[1]{\orgdiv{Department of Space Chemistry and Technology}, \orgname{Heyrovsk{\'y} Institute of the Czech Academy of Sciences}, \orgaddress{\street{Dolej{\v{s}}kova 3}, \city{Prague}, \postcode{182 00}, \country{Czech Republic}}}

\affil[2]{\orgdiv{Department of Theoretical Chemistry}, \orgname{University of Vienna}, \orgaddress{\street{W{\"a}hringer Stra{\ss}e 17}, \city{Vienna}, \postcode{A-1090}, \country{Austria}}}

\affil[3]{\orgdiv{Algorithmic Cheminformatics Group, Faculty of Technology}, \orgname{Bielefeld University}, \orgaddress{\street{Universit{\"a}tsstrasse 25}, \city{Bielefeld}, \postcode{D-33615}, \country{Germany}}}

\affil[4]{\orgdiv{Center for Biotechnology (CeBiTec), Faculty of Biology}, \orgname{Bielefeld University}, \orgaddress{\street{Universit{\"a}tsstrasse 27}, \city{Bielefeld}, \postcode{D-33615}, \country{Germany}}}

\affil[5]{\orgdiv{Department of Mathematics and Computer Science}, \orgname{University of Southern Denmark}, \orgaddress{\street{Campusvej 55}, \city{Odense}, \postcode{5230}, \country{Denmark}}}

\affil[6]{\orgdiv{Bioinformatics Group, Department of Computer Science, and Interdisciplinary Center for Bioinformatics}, \orgname{Universit{\"a}t Leipzig}, \orgaddress{\street{H{\"a}rtelstra{\ss}e 16--18}, \city{Leipzig}, \postcode{D-04107}, \country{Germany}}}

\affil[7]{\orgdiv{German Centre for Integrative Biodiversity Research (iDiv) Halle-Jena-Leipzig \& Competence Center for Scalable Data Services and Solutions Dresden-Leipzig \& School of Embedded and Composite Artificial Intelligence (SECAI) \& Leipzig Research Center for Civilization Diseases}, \orgname{University Leipzig}, \orgaddress{\city{Leipzig}, \postcode{D-04107}, \country{Germany}}}

\affil[8]{\orgname{Max Planck Institute for Mathematics in the Sciences}, \orgaddress{\street{Inselstra{\ss}e 22}, \city{Leipzig}, \postcode{D-04103}, \country{Germany}}}

\affil[9]{\orgdiv{Facultad de Ciencias}, \orgname{Universidad Nacional de Colombia, Sede Bogot{\'a}}, \orgaddress{\street{Ciudad Universitaria}, \city{Bogot{\'a}}, \postcode{COL-111321}, \state{D.C.}, \country{Colombia}}}

\affil[10]{\orgname{Santa Fe Institute}, \orgaddress{\street{1399 Hyde Park Rd.}, \city{Santa Fe}, \state{NM}, \postcode{87501}, \country{USA}}}


\abstract{Pairwise graph ``alignments'', usually defined as maximum common
  subgraphs, have been considered in many application domains, including
  the comparison of molecular structures. Most commonly, the problem is
  formulated as maximum common induced subgraph (MCIS) or maximum common
  edge subgraph (MCES) problems. Sometimes these formulations are
  restricted to connected common subgraphs. Common subgraph problems
  readily generalize to more than two input graphs.  However, this concept
  is fundamentally distinct from the notion of multiple sequence
  alignments.  Abstractly, alignments are defined as super-structures from
  which the individual input objects are recovered via structure-preserving
  projections. While the mathematical construction readily applies to
  common induced subgraphs, it does not fully encompass the common
  edge subgraph problem, which is the variant more commonly employed in
  cheminformatics applications.
	
  In order to consistently define multiple edge-based alignments of graphs
  we represent a graph $G$ as E-graph $(E,\mathscr{V})$, where $E$ is
  the edge set of $G$ and $\mathscr{V}$ is the set of sets of incident
  edges. This representation implicitly encodes the vertices in terms of
  edge data. Alignments of E-graphs are themselves E-graphs that contain
  the input (E-)graphs as subgraphs. Maximum edge subgraphs correspond to
  maximum induced sub-E-graphs, and they can thus be computed using a
  slightly modified version of the VF2 algorithm without the need to switch
  from graphs to line graphs. Furthermore, multiple alignments of
  (E-)graphs yield a natural definition of a consensus graph that is
  ``on average'' present in the input without having to be a subgraph of
  every (and in fact of any) input graph.}

\keywords{maximum common subgraph,
	set systems,
	progressive alignment,
	consensus structure,
	E-graphs,
	graph comparison}

\maketitle

\section{Introduction}\label{sec1}

Graph comparison has been an important topic in computer science and in
many of its application domains. Over the decades, many different
approaches to quantify graph similarity have been proposed. Roughly, they
fall into two classes: \emph{statistical methods}, often referred to as
graph kernels, which evaluate vectors of descriptors \cite{Ghosh:18}, and
\emph{structural methods} aiming at a detailed comparison of the topology
\cite{Neuhaus:07}. Among the latter, graph editing \cite{Sanfeliu:83} asks
for a minimum cost sequence of transformations (usually insertion/deletion
of nodes and/or edges as well as substitutions of vertex and/or edge
labels) that converts one graph into the other. Motivated by quadratic
assignment problems(QAPs), ``graph matching'' or ``graph alignment''
considers 1-1 correspondences (i.e., matchings) between subsets of vertices
and evaluates this partial superposition of graphs using cost functions
that depend on the implied overlap of edges as well as the similarity of
vertex and edge labels \cite{Berg:04,Weskamp:07,Kuchaiev:10,Mernberger:11}.

The concept of alignments used in the context of strings (i.e., DNA, RNA,
or protein sequences) \cite{Morgenstern:99} as well as rooted ordered trees
and forests \cite{Jiang:95,Hoechsmann:04}, is more constrained than graph
matching, however. The key difference is that an alignment is defined as a
super-string or super-tree (carrying multiple labels) from which the input
sequences, trees, or forests are recovered by projection. The notions of
graph matching (sometimes called ``graph alignments'') mentioned above
fit into this paradigm only if matchings are considered not only between
vertices but also explicitly between edges. This formal connection is
further detailed in works such as \cite{Bunke:97,Brun:12}, where it is
shown that optimal graph editing amounts to solving a maximum common
subgraph problem, at least for a large class of cost functions.  In
\cite{Bunke:00} a close connection to the minimum common supergraph is
described. A detailed account of graph alignments in this strict sense can
be found in \cite{GonzalezLaffitte:24a}.

Graph comparisons in chemistry typically consider the structural overlap of
a pair of molecules in order to identify chemical reaction sites, to
compute atom maps for chemical reactions, and to evaluate
structure-activity relationships or the drug-likeness of compounds. Maximum
common subgraph approaches are typically used to identify the
overlapping substructures, see \cite{ehrlich2011,Duesbury:17,duesbury2018}
for reviews.  Different variants of the problem have been considered: the
Maximum Common Induced Subgraph (MCIS) problem asks for the largest
possible number of matched vertices between the two molecules. Although
MCIS is the most natural variant from a mathematical point of view, the
induced subgraph structure is often considered too restrictive for chemical
applications.  The Maximum Common Edge Subgraph (MCES) problem in contrast
maximizes the number of matched edges. Since two incident edges identify a
unique shared vertex, they also imply a matching between vertices (except
in the case of isolated edges, as we shall see below). The MCES is
typically closer to chemist's intuition, e.g.\ in the context of computing
atom-maps \cite{ehrlich2011}.  Both MCIS and MCES as well as the
constrained versions in which the common subgraph must be connected (see
e.g.\ \cite{Vismara:08}) are NP-hard, and their decision versions are
NP-complete \cite{Garey:79}. Common substructures of more than two
  molecules have been addressed by multiple-MCS methods
  \cite{Hariharan:11,Dalke:13} and, more recently, by an exact framework
  that enumerates all (connected) maximum common induced or edge subgraphs
  of multiple labeled graphs by means of recursive modular products and
  clique enumeration \cite{Akbar:27}. These approaches determine
  substructures that are shared by \emph{all} input graphs. We note,
finally that a notion of graph alignment has been used in the translation
of 2D images to molecular graph structures \cite{Oldenhof:22}.  Although
the chemistry literature emphasizes the idea of ``aligning'' two molecules
to identify the common substructure, the notion of an alignment has not
been fully formalized, and multiple alignments of graphs, which also
  account for substructures shared by only some of the inputs, do not seem
to play a role in cheminformatics.

A formal account of alignments as a mathematical structure that is
applicable to very general objects was given recently in
\cite{Berkemer:20b}.  It only requires that the objects to be aligned can
be viewed as sets $X$ of ``atoms'' together with some additional structure
describing some kind of relationships among the ``atoms''. The only formal
requirement is that one can meaningfully obtain sub-objects by restricting
$X$ to a subset $X'\subseteq X$ in a way that relationships among the
``atoms'' are inherited. The usual concepts of string alignments
\cite{Morgenstern:99}, as well as the tree and forest alignments
\cite{Jiang:95,Hoechsmann:04} naturally are of this type, as is a version
of graph alignment in which the allowed matchings between pairs of input
graphs identify the common induced subgraphs. Optimal (pairwise) graph
alignments therefore correspond to solutions of the (weighted) MCIS problem
\cite{Berkemer:20b}. A key advantage of the framework developed in
\cite{Berkemer:20b} is that it not only accounts for pairwise comparisons
but also provides well-defined multiple alignments as well as a natural
way of composing alignments into alignments of alignments.
This form of composability is the key property for employing
generic heuristics to compute multiple graph alignments. Here we
  use the idea of progressive alignments \cite{Feng:87} to
  demonstrate usefulness of composability.

In the setting of cheminformatics, most authors prefer to consider
matchings between bonds (edges of chemical graphs) over matching atoms
(vertices) \cite{ehrlich2011}. In this contribution we therefore develop
the mathematical framework to treat graphs as edge sets with an additional
set-theoretic structure corresponding to the vertices. We call
  these strucutures \emph{E-graphs}.  We note that our approach is
conceptually similar to the use of line graphs as edge-centered description
of graphs proposed e.g.\ in \cite{duesbury2018}, and that E-graphs are
  closely related to the dual hypergraphs of graphs \cite{Berge:89}, see
  Sect.~\ref{sec:egraphs}. However, isomorphism of line graphs implies
isomorphism of the underlying graphs only with the exception $K_3$ and
$K_{1,3}$ \cite{Whitney:32}. Line-graph based methods for the MCES
  problem therefore have to detect and exclude the corresponding
  ``$\Delta Y$ exchanges'' explicitly, as done e.g.\ in RASCAL
  \cite{Raymond:02}. While this is not much of a practical problem in
pairwise comparisons, it complicates the situation in multiple alignments
since the exception would have to be handled in all pairwise comparisons
simultaneously. Here, we introduce E-graphs as an alternative to line
graphs and show that E-graphs are isomorphic if and only if their
corresponding graphs are isomorphic, thus avoiding the problematic issues
of line graphs. Comparisons of E-graphs also do not solve the MCES directly
because of ambiguities arising from isolated edges. More precisely, the
matching of a pair of isolated edges does not determine their relative
orientation, and therefore does not establish an unambiguous correspondence
between their incident vertices. As we shall see below, this ambiguity is
not difficult to handle. The framework presented here makes it
possible to approach the edge-centered graph alignment problem in three
steps that are very similar to approaches for solving the MCIS: First the
maximal common sub-E-graphs are enumerated; this can be done with a
slightly adapted version of the well-known VF2 approach \cite{Cordella:04}.
In the second step, the two E-graphs are ``glued together'' at the matching
edges. Ambiguities arising from isolated matches are resolved at this step,
possibly resulting in more than one alignment for a given maximal common
induced sub-E-graph. General compositional properties of alignments
\cite{Berkemer:20b} can be invoked in the third step to combine pairwise
alignments into a single multiple alignment.

\section{Theory}

\subsection{Abstract Alignments}\label{sec:abstract} 

Alignments can be defined for very general finite structures. In fact, the
only formal pre-requisite is the existence of well-defined subspaces
\cite{Berkemer:20b}. More precisely, let $(X,\mathscr{S})$ be a finite
space, where $X$ denotes the set of points and $\mathscr{S}$
signifies its structure, which may be e.g.\ a partial or total order, a
tree structure, the edges of a directed or undirected graph. We need to be
able to construct a subspace $(X,\mathscr{S})[Y]:=(Y,\mathscr{S}|_{Y})$ by
restricting $(X,\mathscr{S})$ to an arbitrary subset $Y\subseteq X$ in such
a way that $\mathscr{S}|_{Y}$ is uniquely defined. Furthermore, we need
that the construction of subspaces can be composed, i.e., whenever
$Z\subseteq Y\subseteq X$ then we require
$(X,\mathscr{S})[Z]=((X,\mathscr{S})[Y])[Z]$. Intuitively this condition
means that we construct subspaces stepwisely and always obtain the same
result. This condition is not particularly restrictive, in fact it is
satisfied by all the commonly used data structures, including strings,
lists, forests, directed and undirected graphs, as well as all hereditary
graph classes such as planar graphs, cographs, etc. The condition is not
met, say, by the class of graphs with diameter at most $2$ because vertex
deletion may increase the diameter beyond this threshold. The same
structure also pertains to mathematical objects such as topological spaces
or partial orders. We therefore will speak abstractly of \emph{spaces}
$(X_{\alpha},\mathscr{S}_{\alpha})$, where $\alpha$ is taken from some
index set $S$, as the ``input objects''.  From a category theory point of
view, alignments are constructed from the objects of any concrete category
in which the restriction to every subset is a subobject. This includes in
particular all hereditary set systems.  We will only be interested in
finite sets $X_{\alpha}$. The following presentation closely follows
\cite{Berkemer:20b}:

\begin{definition}\label{Def:Alignment}
  An alignment of spaces $(X_{\alpha},\mathscr{S}_{\alpha})$, $\alpha\in
  S$, $|S|\ge 1$ can be defined as a space $(X,\mathscr{S})$ such that
  \begin{description}
  \item[(i)] there is a monomorphism $\mu_{\alpha}:X_{\alpha}\to X$ for every
    $\alpha\in S$;
  \item[(ii)] for every $x\in X$, $\mu^{-1}_{\alpha}(x)\ne\emptyset$ for at
    least one $\alpha\in S$;
  \item[(iii)] The restriction of $(X,\mathscr{S})[\mu_{\alpha}(X_{\alpha})]$
    is isomorphic to $(X_{\alpha},\mathscr{S}_{\alpha})$ with the restriction
    of $\mu_{\alpha}$ to the range $\mu_{\alpha}(X_{\alpha})\subseteq X$
    being an isomorphism. 
  \end{description}
\end{definition}

\begin{figure}[t]
  \begin{center}
    \includegraphics[width=0.6\columnwidth]{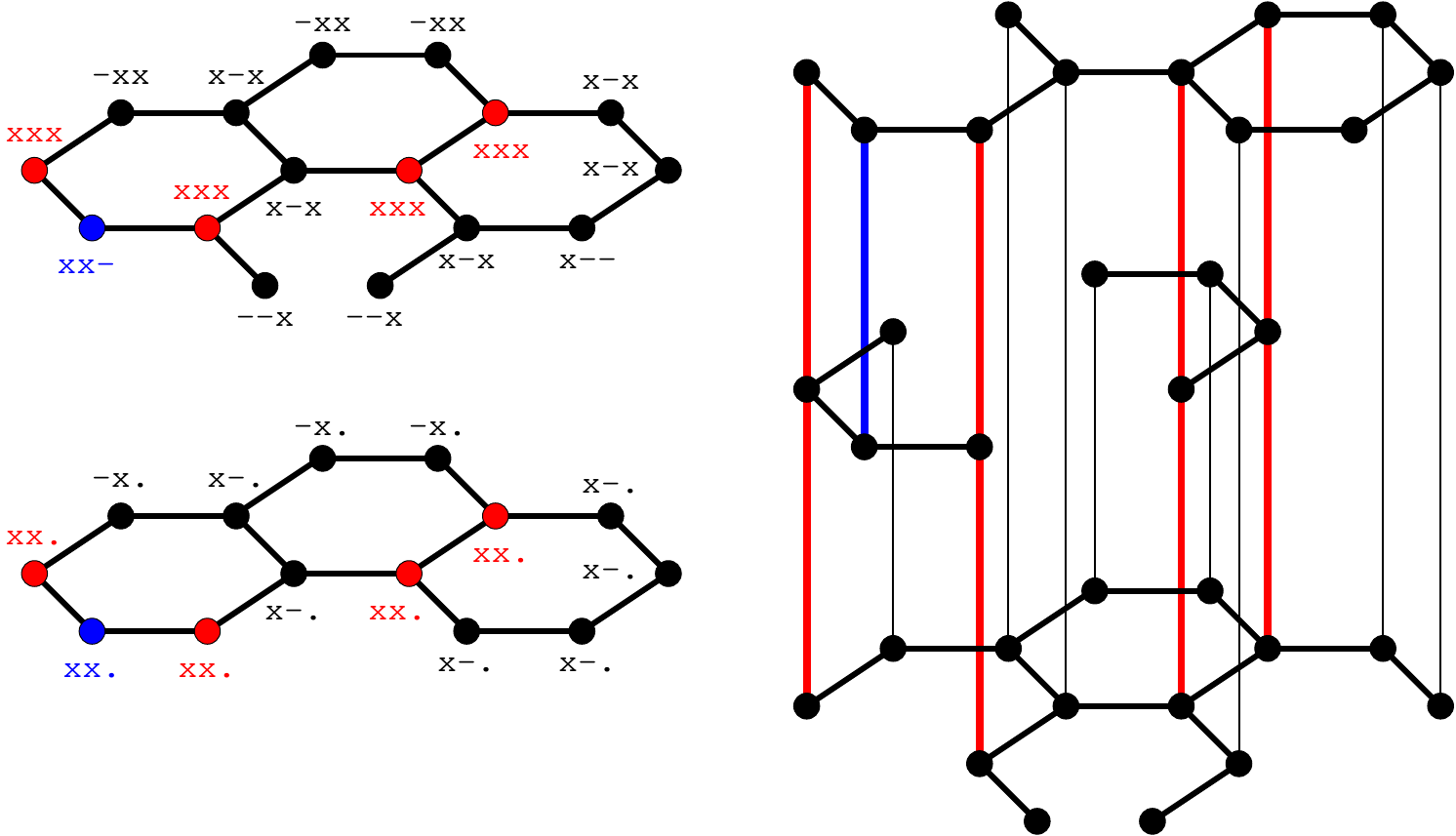}
  \end{center}
  \caption{Vertex-centered alignment of graphs. In the top-left graph,
    every vertex $x$ is labeled with the presence/absence pattern of the
    three rows. Vertices corresponding to match columns $Q_x$ are marked in
    red.  Absent rows in a column correspond to a gap symbol \texttt{-} in
    the graph on the right.  On the left, the three induced subgraphs are
    shown, with vertical lines indicating matches (these are not graph
    edges). Again matches of all three graphs are indicated in red. If only
    the first two graphs are considered, the lower-left alignment remains.
    The matches are now both the vertices shown in red (\texttt{xxx}) and
    the single vertex in blue (\texttt{xx-}) in the three-way
    alignment. The corresponding pairwise alignment consists of all
    vertices with label patterns other than \texttt{--x}, i.e., all but the
    two bottom-most vertices. Moreover, the top-left alignment can be seen
    as an alignment of the bottom-left graph and the graph on the bottom of
    the diagram on the r.h.s.}
  \label{fig:vgr-xmpl}
\end{figure}

It will be convenient to refer to $\alpha$ as \emph{row} of the alignment,
that is common in the context of sequence alignments. For some
$\alpha\ne\beta\in S$, we say that points $x_{\alpha}\in X_{\alpha}$ and
$x_{\beta}\in X_{\beta}$, form a \emph{match} (or are \emph{matched}) if
$\mu_{\alpha}(x_{\alpha})=\mu_{\beta}(x_{\beta})$. Regarding the input
  point sets as pairwise disjoint, the set
$Q_x:=\bigcup_{\alpha\in S}\mu_\alpha^{-1}(\{x\})$ is the \emph{alignment
  column} of $x\in X$.  Note that by construction, the alignment columns
form a partition of the disjoint union $\bigcup_{\alpha\in S} X_{\alpha}$.
For an alignment column $Q_x$ we distinguish whether $\alpha$ is present
(``matched''), i.e., $\mu_{\alpha}^{-1}(x)\ne\emptyset$, or absent
(``deleted''), i.e, $\mu_{\alpha}^{-1}(x)=\emptyset$.  For pairwise
alignments, i.e., $|S|=2$, it therefore makes sense to distinguish
\emph{match columns} $Q_x$ with
$\mu_{\alpha}(x_{\alpha})=\mu_{\beta}(x_{\beta})$ and \emph{in/del
  columns}, where either $\mu_{\alpha}^{-1}(x)=\emptyset$ or
$\mu_{\beta}^{-1}(x)=\emptyset$. In the general case we will call $Q_x$ a
match column if $\mu_{\alpha}^{-1}(x)\ne\emptyset$ for all
$\alpha\in S$. In other words, an alignment column in which all rows are
present, is a match column, see Fig.~\ref{fig:vgr-xmpl}.

A useful property of alignments is that the restriction of
$(X,\mathscr{S})[Y]$ to $Y\coloneqq \bigcup_{\alpha\in S'}
\mu_{\alpha}(X_{\alpha})\subseteq X$ for some subset $S'\subset S$ is an
alignment of spaces $(X_{\alpha},\mathscr{S}_{\alpha})$, where $\alpha \in
S'$ .  It is obtained by simply deleting from $(X,\mathscr{S})$ columns
$Q_x$ with $x\in X\setminus Y$, i.e., the columns in which all present rows
belong to $S\setminus S'$, and thus all rows in $S'$ are
missing. Restricting the resulting alignment to only the rows $\alpha\in
S'$ results in an alignment of the input objects
$(X_{\alpha},\mathscr{S}_{\alpha})$ for $\alpha\in S'$.\footnote{More
  colloquially, this alignment is obtained by first retaining from
  $(X,\mathscr{S})$ on the rows $\alpha\in S'$ and in the result all gap-only
  columns are deleted. Strictly speaking, the intermediate result is not an
  alignment precisely because of the presence of gap-only columns.}

Consequently, a partition of $S=\{S^{(1)},S^{(2)},\dots,S^{(k)}\}$ yields
alignments $(X^{(i)},\mathscr{S}_i)$ for which $\alpha\in S^{(i)}$ and
$(X,\mathscr{S})$ can also be viewed as an alignment of the alignments
$(X^{(i)},\mathscr{S}_i)$ for $1\le i\le k$.  Similarly, the restriction
$(X,\mathscr{S})[Z]$ is an alignment of the restrictions
$(X_{\alpha},\mathscr{S})[\mu_{\alpha}^{-1}(Z)]$ for $\alpha\in S$.  For
details we refer to \cite{Berkemer:20b}.

As an illustrative example, consider the concrete category of finite
simple graphs, considered as a set of vertices with the edges imposed as
additional structure. For graph alignments with no edges between columns having disjoint
row supports, condition (iii) of Definition~\ref{Def:Alignment} can be
expressed in graph-theoretical terms as follows.
\begin{description}
\item[(i)] There is an edge $e\in E$ between $x,y\in X$ in the alignment
  $(X,E)$ if and only if there is $\alpha\in S$ such that
  $\mu^{-1}_\alpha(x)$ and $\mu^{-1}_\alpha(y)$ form an edge in the input
  graph $(X_{\alpha},E_{\alpha})$.
\item[(ii)] If $\mu_{\alpha}(x_{\alpha})=\mu_{\beta}(x_{\beta})$ and
  $\mu_{\alpha}(y_{\alpha})=\mu_{\beta}(y_{\beta})$, where
  $x_{\alpha},y_{\alpha}\in X_{\alpha}$ and
  $x_{\beta},y_{\beta}\in X_{\beta}$,
  then either both
  $x_{\alpha}y_{\alpha}\in E_{\alpha}$ and
  $x_{\beta}y_{\beta}\in E_{\beta}$ or both
  $x_{\alpha}y_{\alpha}\notin E_{\alpha}$ and
  $x_{\beta}y_{\beta}\notin E_{\beta}$.
\end{description}
The subset of vertices in $X_{\alpha}$ and $X_{\beta}$ that are matched
therefore form a \emph{common induced subgraph} of
$(X_{\alpha},E_{\alpha})$ and $(X_{\beta},E_{\beta})$. This remains
true for the vertices that are matched between any subset of input graphs
$(X_{\alpha},E_{\alpha})$, $\alpha\in S'$, for any subset $S'\subseteq S$
comprising at least two input graphs. The columns that consist of matches
only in the graph alignment form a common induced subgraph also in multiple
alignments. For cost functions that independently score mismatches and
insertions/deletions of individual vertices, the problem of computing
optimal pairwise graph alignments is therefore equivalent to computing
maximal common induced subgraphs.  It also involves selecting the subgraph
that maximizes the score from these alignments, see
\cite{Bunke:97,Bunke:00,Brun:12,Berkemer:20b}. Vertex-centered graph
alignments and the construction of progressive multiple alignments of this
type are the subject of \cite{GonzalezLaffitte:24a}.

\subsection{E-graphs} \label{sec:egraphs}

In order to construct alignments in which the edges of a graph (instead of
its vertices) are considered as basic objects, we have to find a
representation of graphs in which (i) edges take on the role of ``points'',
(ii) the vertices of the graphs are described as a set system defined on
this basic point set, and (iii) the resulting set system is hereditary in
the sense that restricting the set of points (i.e., graph edges), specifies
subgraphs in a meaningful manner.  We cannot consider graphs with isolated
vertices because isolated vertices -- by definition -- are not
associated with any graph edge, and therefore cannot be expressed as a set
of graph edges. As we shall see below, furthermore, isolated graph edges
introduce additional complications. In order to motivate our construction,
we begin with a simple observation on graphs without isolated vertices and
edges.
\begin{lemma}
  Let $G$ be a graph without isolated vertices and isolated edges. Then
  there is a 1-1 correspondence between the vertices $v$ and sets
  $\Upsilon(v):=\{e\in E| v\in e\}$. 
  \label{lem:Upsilon}
\end{lemma}
\begin{proof}
  By construction, $\Upsilon(v)$ is uniquely determined by $v$. It thus
  suffices to show that, conversely, a set $\Upsilon$ identifies $v$.
  First assume $|\Upsilon|>1$. Since $\Upsilon$ by assumption is a set of
  edges meeting a single vertex $v$, we have
  $\bigcap_{e\in \Upsilon}e =\{v\}$. Now suppose that $|\Upsilon|=1$, i.e.,
  it consists of a single edge $e=uv\in E$. By assumption $G$ contains no
  isolated edges, thus $e$ has one end, say $v$, that has vertex degree
  larger than $1$, and thus there is a set $\Upsilon'$ with $e\in\Upsilon'$
  and $|\Upsilon'|>1$. Then $u$ must have degree $1$ and thus is incident
  to $e$ only.
\end{proof}

Let $G$ be a graph with vertex set $V$ and edge set $E$. Furthermore, let
$V_3(G)$ be the set of all vertices in $G$ that reside in a connected
component with at least 3 vertices or, equivalently, at least 2 edges.
That is, $V=V_3(G)$ if and only if $G$ has neither isolated vertices nor
isolated edges.  For $V=V_3(G)$, consider the set system
$(E(G),\mathscr{V}_G)$ with
$\mathscr{V}_G\coloneqq\{ \Upsilon(v)| v\in V(G)\}$. It follows immediately
from the construction that (i) for distinct graph vertices
$u,v\in V$, we have $|\Upsilon(u)\cap \Upsilon(v)|=1$ if $u$ and $v$ are
adjacent and $|\Upsilon(u)\cap \Upsilon(v)|=0$ otherwise; and (ii) every
edge $e\in E(G)$ is contained in exactly two distinct sets
$\Upsilon,\Upsilon'\in\mathscr{V}_G$.

\begin{lemma}
  \label{lem:V3}
  Let $E$ be a set and $\mathscr{V}\subseteq 2^E$ such that (i) $\Upsilon$
  is nonempty, for every $\Upsilon\in \mathscr{V}$, (ii)
  $|\Upsilon\cap\Upsilon'|\le 1$ for any two distinct
  $\Upsilon,\Upsilon'\in\mathscr{V}$ and (iii) every $e\in E$ appears in
  exactly two elements of $\mathscr{V}$.  Then the graph $G$ with vertex
  set $\mathscr{V}$ and edges $\{\Upsilon,\Upsilon'\} \in E(G)$ if and only
  if $|\Upsilon\cap \Upsilon'|=1$ for distinct
  $\Upsilon,\Upsilon'\in \mathscr{V}$ has no isolated vertices or edges.
\end{lemma}
\begin{proof}
  Clearly, $G=(\mathscr{V},E(G))$ is well-defined and indeed a graph.  By
  assumption (i), there is $e\in\Upsilon$, and by (iii) $e$ is also
  contained in a second set $\Upsilon'\ne\Upsilon$. By definition of
  $E(G)$, therefore, $\Upsilon$ is incident with an edge, and thus $G$
  contains no isolated vertices. Since $\Upsilon'\ne\Upsilon$, at least one
  of them, say $\Upsilon'$, has cardinality larger than $1$ and thus also
  contains a second edge $e'\ne e$. Since $\Upsilon\cap \Upsilon'=\{e\}$,
  we have $e'\notin\Upsilon$ and thus, invoking (iii), there is a
  third set $\Upsilon''\notin\{\Upsilon,\Upsilon'\}$ such that
  $\Upsilon'\cap\Upsilon''=\{e'\}$.  Thus $\Upsilon$ is part of a connected
  component in $G$ that contains at least three distinct vertices
  $\Upsilon,\Upsilon',\Upsilon''\in\mathscr{V}$ and the two connecting
  edges $e$ and $e'$.
\end{proof}

Lemmas \ref{lem:Upsilon} and \ref{lem:V3} imply that there is a 1-1
correspondence between set systems $(E,\mathscr{V})$ satisfying the
conditions in Lemma~\ref{lem:V3} and graphs $G=(V,E)$ without isolated
vertices and edges. Indeed, starting from $G=(V,E)$ we see $\Upsilon(u)$
and $\Upsilon(v)$ are adjacent in $(\mathscr{V},E(G))$ if and only if
$\Upsilon(u)\cap\Upsilon(v)=\{e\}$, which is the case if and only if $e$
was incident to both $u$ and $v$ in $G$. Hence $G=(V,E)$ and the graph
$(\mathscr{V},E(G))$ are isomorphic. In particular, there is a 1-1
correspondence between $E$, the set of points in the E-graph, and the set
$E(G)$ in the corresponding graph.

\begin{figure}[t]
	\begin{center}
		\includegraphics[width=0.45\columnwidth]{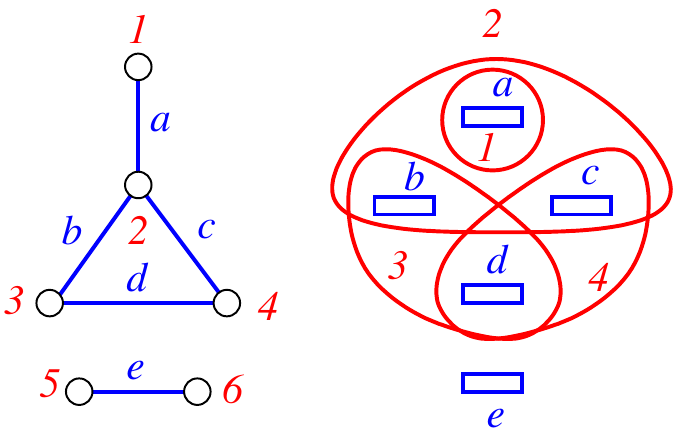}
	\end{center}
	\caption{The graph $G$ (left) with vertex set $V=\{1,2,3,4,5,6\}$ and
		edge set $E=\{a,b,c,d,e\}$ has an isolated edge $e=\{5,6\}$. In the
		corresponding E-graph the graph vertices $5$ and $6$ that form the
		graph edge $e$ are not represented because $e$ is an isolated point in
		the E-graph.  The other four graph vertices are specified by the sets
		of points in the E-graph: $1=\{a\}$, $2=\{a,b,c\}$, $3=\{b,d\}$, and
		$4=\{c,d\}$.  These are exactly the (pairwise distinct) sets of graph
		edges that are incident with a graph vertex. The isolated point $e$, on
		the other hand, does not appear in any of incidence sets.}
	\label{fig:egraph}
\end{figure}

As noted above, isolated vertices in $G$ are irrelevant because we will be
concerned with alignments that consider the edges as points. Isolated edge,
however, need to be considered. While in connected components with at least
two edges the vertices are identified uniquely by the edges to which they
are incident, this is not the case for an isolated edge $e=uv$. In this
case, the edges of $G$ do not provide any information that could be used to
distinguish the endpoints $u$ and $v$ of $e$. This suggests to add isolated
edges in $G$ as isolated points to $(E,\mathscr{V})$. That is, $e\in E$ is
isolated if $e\notin\Upsilon$ for any $\Upsilon\in\mathscr{V}$.
\begin{definition}
	\label{Def:E-graph}
	An \emph{E-graph} is a pair $(E,\mathscr{V})$ with $\mathscr{V}\subseteq
	2^E$ and $\emptyset\notin\mathscr{V}$, such that (i) every $e\in E$ is
	contained either in exactly two or in none of the elements of
	$\mathscr{V}$ and (ii) $|\Upsilon\cap\Upsilon'|\le 1$ for all distinct
	$\Upsilon,\Upsilon'\in\mathscr{V}$.
\end{definition}
The relationship of a graph and its corresponding E-graph representation is
shown in Fig.~\ref{fig:egraph}. Importantly, given a graph $G$, its
corresponding E-graph is uniquely defined. The converse is not true for
E-graphs with isolated points: the corresponding graph is determined only
up to the identity of the vertices incident with the isolated edges.
More precisely, let $E^{\circ}\coloneqq\bigcup_{\Upsilon\in\mathscr{V}}
\Upsilon$ denote the set of non-isolated points of an E-graph
$(E,\mathscr{V})$. Then $(E^{\circ},\mathscr{V})$ satisfies the conditions
of Lemma~\ref{lem:V3} and thus corresponds to a unique graph without
isolated vertices and edges, while every isolated point
$e\in E\setminus E^{\circ}$ corresponds to a $K_2$ component. Lemmas
\ref{lem:Upsilon} and \ref{lem:V3} therefore extend to E-graphs with
isolated points in the following form: E-graphs are in 1-1 correspondence
with graphs without isolated vertices, considered up to the identity of
the endpoints of their isolated edges.

As simple examples consider the triangle $K_3$ and the claw $K_{1,3}$,
both with edge set $E=\{a,b,c\}$. The E-graph of $K_3$ is
$(E,\{\{a,b\},\{b,c\},\{a,c\}\})$, while the E-graph of $K_{1,3}$ is
$(E,\{\{a,b,c\},\{a\},\{b\},\{c\}\})$. The path $P_4$ with consecutive
edges $a,b,c$ corresponds to $(E,\{\{a\},\{a,b\},\{b,c\},\{c\}\})$, and the
disjoint union of a path with edges $a,b$ and a single edge $c$ corresponds
to $(E,\{\{a\},\{a,b\},\{b\}\})$, in which $c$ is an isolated point.

E-graphs are a slight variation on the intersection representation of
graphs \cite{Erdos66, McKee99}, where the set $\Upsilon(v):=\{e\in E| v\in
e\}$ is assigned to each vertex $v\in V(G)$. Then, as there is no multiple
edge in $G$, two vertices are adjacent if and only if their corresponding
sets intersect in exactly one element. If $G$ has no isolated vertex then
all assigned sets are non-empty. Moreover if there is no isolated edge,
then distinct vertices are represented by distinct sets of labels. In
E-graphs, the sets $\Upsilon\in\mathscr{V}$ identify exactly the vertices
incident with non-isolated edges, while vertices incident with isolated
edges do not have a representation at all.  In E-graphs it is therefore not
necessary to keep track of the assignment $v\mapsto\Upsilon(v)$.

In an alternative interpretation, E-graphs $(E,\mathscr{V})$ form a
particular class of hypergraphs.  In particular, two distinct hyperedges
$\Upsilon,\Upsilon'$ share at most one vertex, thus $(E,\mathscr{V})$ is a
\emph{linear} hypergraph.  This picture is useful e.g.\ to define
connectivity in E-graphs. A hyperpath in $(E,\mathscr{V})$ is an
alternating sequence of points $e_i\in E$ and sets
$\Upsilon_i\in\mathscr{V}$ such that $e_i,e_{i+1}\in \Upsilon_i$. The
E-graph $(E,\mathscr{V})$ is connected if there is a hyperpath connecting
any two points $e,e'\in E$. Any isolated point is considered to be
connected. Using Lemmas \ref{lem:Upsilon} and \ref{lem:V3} it is easy to
verify that a graph $G=(V, E)$ is connected if and only if its
corresponding E-graph $(E,\mathscr{V})$ is connected.

For a graph $G$ without isolated vertices and edges, $(E,\mathscr{V}_G)$
is precisely the \emph{dual hypergraph} $G^*$ of $G$ \cite{Berge:89}: its
vertices are the edges of $G$ and its hyperedges are the vertex stars
$\Upsilon(v)$, i.e., its incidence matrix is the transpose of the
incidence matrix of $G$. Linearity, the connectivity statement above, and
the fact that $G$ is recovered from $G^*$ (Lemmas \ref{lem:Upsilon} and
\ref{lem:V3}) are therefore inherited from standard hypergraph theory. The
difference between E-graphs and dual hypergraphs lies in the treatment of
isolated edges: the dual of an isolated edge $uv$ consists of a single
vertex contained in the two indistinguishable singleton hyperedges
$\Upsilon(u)=\Upsilon(v)=\{uv\}$, whereas Def.~\ref{Def:E-graph}
represents it as an isolated point that is not contained in any set. As
we shall see in Lemma~\ref{lem:sub-E-graph}, it is this convention that
makes the class of E-graphs closed under restriction to arbitrary subsets
of points, and thus suitable for the alignment framework of Sect.~\ref{sec:abstract}.
	
Recall that two graphs $G=(V,E)$ and $H=(W,F)$ are isomorphic if there is a
bijection $\psi: V\to W$ such that $uv\in E$ if and only if
$\psi(u)\psi(v)\in F$. Analogously, E-graphs are isomorphic if there is a
bijection between their points that preserves the hyperedges. More
formally, we have:
\begin{definition}
	Two E-graphs $\tilde G=(E,\mathscr{V})$ and $\tilde H=(F,\mathscr{W})$
	are isomorphic if there is a bijection $\varphi: E\to F$ such that
	$\Upsilon\in\mathscr{V}$ if and only if
	$\varphi(\Upsilon)\in\mathscr{W}$.
\end{definition}

\begin{theorem}
	\label{thm:isomorph}
	Let $(V,E)$ and $(W,F)$ be two graphs without isolated vertices and let
	$(E,\mathscr{V})$ and $(F,\mathscr{W})$ be the corresponding
	E-graphs. Then $(V,E)$ and $(W,F)$ are isomorphic if and only if
	$(E,\mathscr{V})$ and $(F,\mathscr{W})$ are isomorphic.
\end{theorem}
\begin{proof}
	We first note that every graph isomorphism is a disjoint union of
	isomorphisms between connected components. Furthermore, all $K_2$
	components (isolated edges) are isomorphic. Similarly, every E-graph
	isomorphism is the disjoint union of E-graph isomorphisms between
	connected components. Here, all isolated points are isomorphic. It
	follows from the definition of E-graphs that there is a 1-1
	correspondence of the isolated edges in graph and the isolated points in
	the corresponding E-graph. Hence $(V,E)$ and $(W,F)$ are isomorphic if
	and only if they have the same number of isolated edges and there is a
	bijection between the connected components with at least three vertices
	such that corresponding connected components are isomorphic. In the same
	manner, $(E,\mathscr{V})$ and $(F,\mathscr{W})$ are isomorphic, if they
	have the same number of isolated points and there is a bijection between
	the connected components of $(E,\mathscr{V})$ and $(F,\mathscr{W})$ with
	at least two points such that corresponding components are isomorphic
	E-graphs.  It suffices, therefore, to show that the statement is true for
	connected graphs with at least three vertices and connected E-graphs with
	at least two points. Lemmas \ref{lem:Upsilon} and \ref{lem:V3} establish
	a 1-1 correspondence between such graphs and E-graphs. 
	
	Suppose $G=(V,E)$ and $H=(W,F)$ are isomorphic and let $\psi:V\to W$ be
	an isomorphism. Then $\Upsilon_G(v)=\{vu\in E| u\in V\}$ and
	$\Upsilon_H(w)=\{wy\in F| y\in W\}$. Writing, as common in graph theory
	$\psi(uv)\coloneqq\psi(\{u,v\})$, for $w=\psi(v)$ we get
	$\Upsilon_H(\psi(v))=\{\psi(v)\psi(u)\in F| \psi(u)\in W\}=
	\{\psi(v)\psi(u)\in F| u\in V\} = \{\psi(vu)\in F| u\in V\} = \{\psi(uv)|
	uv\in E, u\in V\} = \psi(\Upsilon_G(v))$.  Thus every hyperedge
	$\Upsilon_H(w)$ is the image of a hyperedge $\Upsilon_G(v)$ under $\psi$.
	Every graph isomorphism $\psi$ from $G$ to $H$ therefore induces an
	E-graph isomorphism on the corresponding E-graphs.
	
	Conversely, suppose $(E,\mathscr{V})$ and $(F,\mathscr{W})$ are
	isomorphic E-graphs and let $\varphi$ be an E-graph isomorphism. Note
	that $\varphi$ is, in particular, also a bijection
	$\varphi:\mathscr{V}\to\mathscr{W}$. Let $(V,E')$ and $(W,F')$ be the
	unique graphs corresponding to $(E,\mathscr{V})$ and
	$(F,\mathscr{W})$. In particular, there is a 1-1 correspondence between
	the sets in the set systems $\mathscr{V}$ and $\mathscr{W}$ and the
	vertex sets $V$ and $W$, respectively. Let $\Upsilon_u\in\mathscr{V}$ be
	the set associated with $u\in V$ and $\Gamma_w\in\mathscr{W}$ be the
	set associated with $w\in W$ and set $\psi(u)\eqqcolon w$ if
	$\varphi(\Upsilon_u)=\Gamma_w$. By construction we have $uv\in E'$ if and
	only if $|\Upsilon_u\cap\Upsilon_v|=1$ and $xy\in F'$ if and only if
	$|\Gamma_x\cap \Gamma_y|=1$.  Since $\varphi$ is a bijection, we have
	$|\varphi(\Upsilon_u)\cap\varphi(\Upsilon_v)|=|\Gamma_{\psi(u)}\cap
	\Gamma_{\psi(v)}|$ and thus $uv\in E'$ if and only if $\psi(u)\psi(v)\in
	F'$. Thus the bijection $\psi: V\to W$ is a graph isomorphism and hence
	$(V,E')$ and $(W,F')$ are isomorphic.
\end{proof}

Hassler Whitney's \emph{line graph theorem} states that two connected graphs
distinct from $K_3$ and $K_{1,3}$ are isomorphic if and only if their line
graphs are isomorphic \cite{Whitney:32,Hemminger:72}. The line graph
$\mathcal{L}[G]$ of a graph $G=(V,E)$ has $E$ as its vertex set and $ef$ is
an edge in $\mathcal{L}[G]$ if $|e\cap f|=1$, i.e., if the edges $e$ and
$f$ share a single vertex in $G$.  The line graph theorem serves as
justification to compute the maximum common induced subgraph of the line
graphs $\mathcal{L}[G]$ and $\mathcal{L}[H]$ instead of computing the
maximum common edge subgraph of $G$ and $H$ directly. This approximation
has been used extensively in particular in application to chemistry
\cite{ehrlich2011,Duesbury:17,duesbury2018}; the exceptional case
manifests itself in the form of so-called $\Delta Y$ exchanges, i.e.,
common induced subgraphs of the line graphs that map the three edges of a
triangle onto the three edges of a claw, which have to be detected and
removed explicitly, as done e.g.\ in RASCAL \cite{Raymond:02}. A formal
treatment of this issue was given recently in \cite{Akbar:27}, where
cliques of the modular product of (labeled) line graphs are called
$\Delta$-Y admissible if every triangle of the clique projects to
triangles in all factors or to claws in all factors, and where it is
shown that the maximum $\Delta$-Y admissible cliques correspond exactly
to the maximum common edge subgraphs of the input graphs.
Thm.~\ref{thm:isomorph} can be
seen as an analogue of Whitney's result for E-graphs instead of line
graphs. Clearly, E-graphs have the advantage that the equivalence holds
without exceptions. E-graphs are of course distinct from line
	graphs: both $K_3$ and $K_{1,3}$ have $K_3$ as their line graph, whereas
	their E-graphs, given above, are non-isomorphic. We shall return to this
	example in Sect.~\ref{sec:egraphalign}.

In order for alignments of E-graphs to be well-defined, we need to show
that E-graphs have well-defined induced sub-E-graphs that are uniquely
defined by restricting the point set.
\begin{definition}
	\label{Def:sub-E-graph}
	Let $G=(E,\mathscr{V})$ be an E-graph, let $F\subseteq E$, and construct
	$H=(E,\mathscr{V})[F] =:(F,\mathscr{V}_F)$ by setting $\Upsilon\cap F\in
	\mathscr{V}_F$ if and only if either (i) $|\Upsilon\cap F|\ge 2$ or (ii)
	$\Upsilon\cap F=\{e\}$ and there is $\Upsilon'\in\mathscr{V}$ with
	$|\Upsilon'\cap F|\ge 2$ and $e\in \Upsilon'$. We call
	$(E,\mathscr{V})[F]$ the \emph{sub-E-graph} of $(E,\mathscr{V})$
	\emph{induced by $F$}.
\end{definition}
We note that the 1-1 correspondence of vertices $v\in V$ and sets
$\Upsilon_v\in\mathscr{V}$ is confined to E-graphs without isolated points.
In fact, isolated points do not correspond to \emph{any} set in
$\mathscr{V}$. Thus a set $\Upsilon$ may, but it is not required to,
entirely disappear from $\mathscr{V}$ if $\Upsilon\cap F=\{e\}$. This
depends on whether $e$ is an isolated point in the sub-E-graph
$(F,\mathscr{V}_F)$.
For the E-graph of Fig.~\ref{fig:egraph}, for instance, the
restriction to $F=\{a,b,d\}$ yields the sets $\{a,b\}$, $\{b,d\}$, $\{a\}$,
and $\{d\}$, i.e., the E-graph of the path with consecutive edges $a,b,d$;
the singletons $\{a\}=\Upsilon(1)\cap F$ and $\{d\}=\Upsilon(4)\cap F$ are
retained by condition (ii). The restriction to $F=\{a,e\}$, on the other
hand, yields $\mathscr{V}_F=\emptyset$, since $\Upsilon(1)\cap F=
\Upsilon(2)\cap F=\{a\}$ and no set meets $F$ in more than one point. Both
$a$ and $e$ thus become isolated points, in accordance with the fact that
$a$ is an isolated edge of the subgraph of $G$ formed by the edges $a$ and
$e$.

\begin{lemma}
	Let $(E,\mathscr{V})$ be an E-graph. Then
	$(E,\mathscr{V})[F]$ is an E-graph for all $F\subseteq E$. Moreover,
	the following statements are true:
	\begin{itemize}
		\item[(i)]   $(E,\mathscr{V})[E]=(E,\mathscr{V})$.
		\item[(ii)]  $(E,\mathscr{V})[\{e\}] = (\{e\},\emptyset)$.
		\item[(iii)] $(E,\mathscr{V})[F]=((E,\mathscr{V})[E'])[F]$ for all
		$F\subseteq E'\subseteq E$.
	\end{itemize}
	\label{lem:sub-E-graph}
\end{lemma}
\begin{proof}
	The first alternative in the construction of $H$, the sub-E-graph of
	$(E,\mathscr{V})$ induced by $F$, retains all ``vertices'' $\Upsilon$
	with a degree of least two. The second condition retains all vertices of
	degree 1 if and only if they are not located in isolated edges. Thus
	sub-E-graphs suppress all vertices that have no incident edges and all
	vertices that delimit isolated edges in $H$. Thus $H$ is an
	E-graph. Statement (i) follows from $\Upsilon\cap E=\Upsilon$ and the
	definition of E-graphs. In Statement (ii), $|\Upsilon\cap \{e\}|\le 1$
	and thus $(\Upsilon\cap \{e\})\notin \mathscr{V}_{\{e\}}$ for all
	$\Upsilon\in\mathscr{V}$.  To see Statement (iii), write
	$(E',\mathscr{V}_{E'})\coloneqq(E,\mathscr{V})[E']$ and note that
	$(\Upsilon\cap E')\cap F=\Upsilon\cap F$ for all $F\subseteq E'\subseteq
	E$. If $|\Upsilon\cap F|\ge 2$, then $|\Upsilon\cap E'|\ge 2$, hence
	$\Upsilon\cap E'\in\mathscr{V}_{E'}$ by condition (i) of
	Def.~\ref{Def:sub-E-graph}, and $\Upsilon\cap F$ is retained in
	$((E,\mathscr{V})[E'])[F]$, again by condition (i). If $\Upsilon\cap
	F=\{e\}$ is retained in $(E,\mathscr{V})[F]$ by condition (ii), there is
	$\Upsilon'\in\mathscr{V}$ with $e\in\Upsilon'$ and $|\Upsilon'\cap F|\ge
	2$. Then $|\Upsilon'\cap E'|\ge 2$, so that $\Upsilon'\cap
	E'\in\mathscr{V}_{E'}$, and $\Upsilon\cap E'\in\mathscr{V}_{E'}$ because
	it either has at least two elements or equals $\{e\}$ with
	$e\in\Upsilon'\cap E'$. Hence $\{e\}=(\Upsilon\cap E')\cap F$ is retained
	in $((E,\mathscr{V})[E'])[F]$ by condition (ii), applied to
	$\Upsilon'\cap E'$. Conversely, every set retained in
	$((E,\mathscr{V})[E'])[F]$ is of the form $(\Upsilon\cap E')\cap
	F=\Upsilon\cap F$ and satisfies condition (i) or (ii) with respect to
	$(E,\mathscr{V})$ and $F$, since $|(\Upsilon'\cap E')\cap F|\ge 2$
	implies $|\Upsilon'\cap F|\ge 2$. 
	Thus $\mathscr{V}_F$ is indeed independent of the choice of $E'$
	whenever $F\subseteq E'\subseteq E$.
\end{proof}

	Lemma~\ref{lem:sub-E-graph} also makes precise the relationship between
	induced sub-E-graphs and edge subgraphs. For a graph $G=(V,E)$ and
	$E'\subseteq E$ we write $G\langle E'\rangle$ for the subgraph of $G$
	formed by the edges in $E'$ and their endpoints. Common subgraphs of this
	form are the objects of the MCES problem.
	\begin{proposition}\label{prop:MCES}
		Let $G=(V,E)$ and $H=(W,F)$ be graphs without isolated vertices, and let
		$(E,\mathscr{V})$ and $(F,\mathscr{W})$ be the corresponding
		E-graphs. Then, for every $E'\subseteq E$, the sub-E-graph
		$(E,\mathscr{V})[E']$ is the E-graph of $G\langle E'\rangle$.
		Consequently, for $E'\subseteq E$ and $F'\subseteq F$, the subgraphs
		$G\langle E'\rangle$ and $H\langle F'\rangle$ are isomorphic if and only
		if the sub-E-graphs $(E,\mathscr{V})[E']$ and $(F,\mathscr{W})[F']$ are
		isomorphic. In particular, the common edge subgraphs of $G$ and $H$ with
		$k$ edges correspond to the common induced sub-E-graphs of
		$(E,\mathscr{V})$ and $(F,\mathscr{W})$ with $k$ points, and maximum
		common edge subgraphs correspond to maximum common induced sub-E-graphs.
	\end{proposition}
	\begin{proof}
		Let $v$ be a vertex of $G\langle E'\rangle$, i.e., an endpoint of an edge
		in $E'$. Then $\Upsilon(v)\cap E'=\{e\in E'| v\in e\}$ is the set of
		edges of $G\langle E'\rangle$ incident with $v$. By
		Def.~\ref{Def:sub-E-graph}, $\Upsilon(v)\cap E'\in\mathscr{V}_{E'}$ if
		and only if $v$ has degree at least $2$ in $G\langle E'\rangle$, or $v$
		has degree $1$ and the other endpoint of its unique edge $e\in E'$ has
		degree at least $2$ in $G\langle E'\rangle$; that is, if and only if $v$
		is not an endpoint of an isolated edge of $G\langle E'\rangle$. Vertices
		of $G$ that are not incident with an edge in $E'$ contribute
		$\Upsilon(v)\cap E'=\emptyset\notin\mathscr{V}_{E'}$. Since $G\langle
		E'\rangle$ has no isolated vertices by construction, $(E',\mathscr{V}_{E'})$
		is exactly its E-graph in the sense of Def.~\ref{Def:E-graph}. The second
		statement now follows from Thm.~\ref{thm:isomorph} applied to $G\langle
		E'\rangle$ and $H\langle F'\rangle$. The third statement is a
		reformulation of the second one, since the subgraphs $G\langle E'\rangle$
		with $|E'|=k$ are precisely the subgraphs of $G$ with $k$ edges and
		without isolated vertices.
	\end{proof}
	We emphasize that an isomorphism of sub-E-graphs determines the
	corresponding isomorphism of edge subgraphs uniquely on all components
	with at least two edges, while for every isolated edge of the common
	subgraph both correspondences of its endpoints are possible, cf.\ the
	proof of Thm.~\ref{thm:isomorph}. This ambiguity reappears in
	Sect.~\ref{sec:pairwise} as the choice of an \emph{orientation}.
	Prop.~\ref{prop:MCES} is the E-graph counterpart of the $\Delta$-Y
	disambiguation theorem for line graphs \cite[Thm.~5]{Akbar:27}, which
	establishes the correspondence between common edge subgraphs and
	$\Delta$-Y admissible cliques of the modular product of the line
	graphs. In the E-graph setting no admissibility condition on triangles is
	required, because the sets of an E-graph retain the identity of the
	vertices and thus distinguish a triangle from a claw; the price is that
	isomorphisms of sub-E-graphs have to be tested on set systems rather than
	on graphs, which is what the set-matches of
	Sect.~\ref{sec:egraphalign} accomplish.

\subsection{Alignments of E-graphs}\label{sec:egraphalign}

Lemma~\ref{lem:sub-E-graph} establishes that E-graphs
(Def.~\ref{Def:E-graph}) and their sub-E-graphs
(Def.~\ref{Def:sub-E-graph}) satisfy the sufficient condition for
well-defined alignments \cite{Berkemer:20b}. For later reference, we
record the following.
\begin{fact}\label{Def:Align_of_E-graphs}
	The E-graph $(E,\mathscr{V})$ is an alignment of the E-graphs
	$(E_{\alpha},\mathscr{V}_{\alpha})$ with $\alpha\in S$,
	$|S|\ge 1$, if
	\begin{description}
		\item[(i)] there is a monomorphism $\mu_{\alpha}:E_{\alpha}\to E$ for every
		$\alpha\in S$;
		\item[(ii)] for every $x\in E$, $\mu^{-1}_\alpha(x)\ne\emptyset$ for at
		least one $\alpha\in S$;
		\item[(iii)] The sub-E-graph $(E,\mathscr{V})[\mu_{\alpha}(E_{\alpha})]$ is
		isomorphic to $(E_{\alpha},\mathscr{V}_{\alpha})$, such that the
		restriction of $\mu_{\alpha}$ to
		$(E,\mathscr{V})[\mu_{\alpha}(E_{\alpha})]$ is an isomorphism.
	\end{description}
\end{fact}

For simplicity we first consider the pairwise case, i.e., $|S|=2$, and
let $M\subseteq E$ be a set of matches, i.e.,
\begin{equation}
	M=\{x\in E\ |\ \mu_1^{-1}(x)\ne\emptyset \text{ and }
	\mu_2^{-1}(x)\ne\emptyset\}\,.
\end{equation}
We write $M_1=\mu_1^{-1}(M)$ and $M_2=\mu_2^{-1}(M)$ for the sets of
matched points in the E-graphs $(E_{1},\mathscr{V}_{1})$ and
$(E_{2},\mathscr{V}_{2})$, respectively. 
\begin{lemma}\label{lem:matchsub}
	Let $(E,\mathscr{V})$ be an alignment of two E-graphs
	$G_1=(E_1,\mathscr{V}_1)$ and $G_2=(E_2,\mathscr{V}_2)$ with a set of
	matches $M\subseteq E$. Then $(E,\mathscr{V})[M]$ is isomorphic to both
	$(E_1,\mathscr{V}_1)[M_1]$ and $(E_2,\mathscr{V}_2)[M_2]$. 
\end{lemma}
\begin{proof}
	It suffices to note that $(E,\mathscr{V})[\mu_1(E_1)]$ and
	$(E,\mathscr{V})[\mu_2(E_2)]$ are isomorphic to $(E_1,\mathscr{V}_1)$ and
	$(E_2,\mathscr{V}_2)$, respectively. Since the restriction of $\mu_1$ to
	$M_1$ and $\mu_2$ to $M_2$ are bijections, we obtain
	$(E,\mathscr{V})[M] \simeq (E_1,\mathscr{V}_1)[M_1]$ and
	$(E,\mathscr{V})[M] \simeq (E_2,\mathscr{V}_2)[M_2]$.
\end{proof}
In other words, every set of matches and, equivalently, every set of
match columns in $(E,\mathscr{V})$ defines a common induced sub-E-graph
of $(E_1,\mathscr{V}_1)$ and $(E_2,\mathscr{V}_2)$. This observation
generalizes immediately to match columns in larger alignments, i.e., to
sets of matches $M$ such that for all $x\in M$ and all $\alpha\in
S'\subseteq S$ it holds that $\mu^{-1}_{\alpha}(x)\ne\emptyset$.  The
set of match columns in the restriction of an E-graph alignment to a
subset $S'$ of $S$ therefore forms a common induced sub-E-graph of the
E-graphs $(E_{\alpha},\mathscr{V}_{\alpha})$ with $\alpha\in S'$.

Consider three E-graphs $(F,\mathscr{V})$, $(E_1,\mathscr{V}_1)$, and
$(E_2,\mathscr{V}_2)$ and suppose $(F,\mathscr{V})$ is (isomorphic to) a
sub-E-graph of both $(E_1,\mathscr{V}_1)$ and $(E_2,\mathscr{V}_2)$,
i.e., there are injective maps $\mu_1:F\to E_1$ and $\mu_2:F\to E_2$ such
that the restriction $(E_1,\mathscr{V}_1)[\mu_1(F)]\simeq
(F,\mathscr{V})$ and $(E_2,\mathscr{V}_2)[\mu_2(F)]\simeq
(F,\mathscr{V})$. We can think of $(F,\mathscr{V})$ as an alignment of
$(E_1,\mathscr{V}_1)[\mu_1(F)]$ and $(E_2,\mathscr{V}_2)[\mu_2(F)]$ that
consists only of matches. Naturally, we ask whether there is an alignment
of $(E_1,\mathscr{V}_1)$, and $(E_2,\mathscr{V}_2)$ that has exactly this
set of matches.

\begin{figure}
	\begin{center}
		\includegraphics[width=0.7\columnwidth]{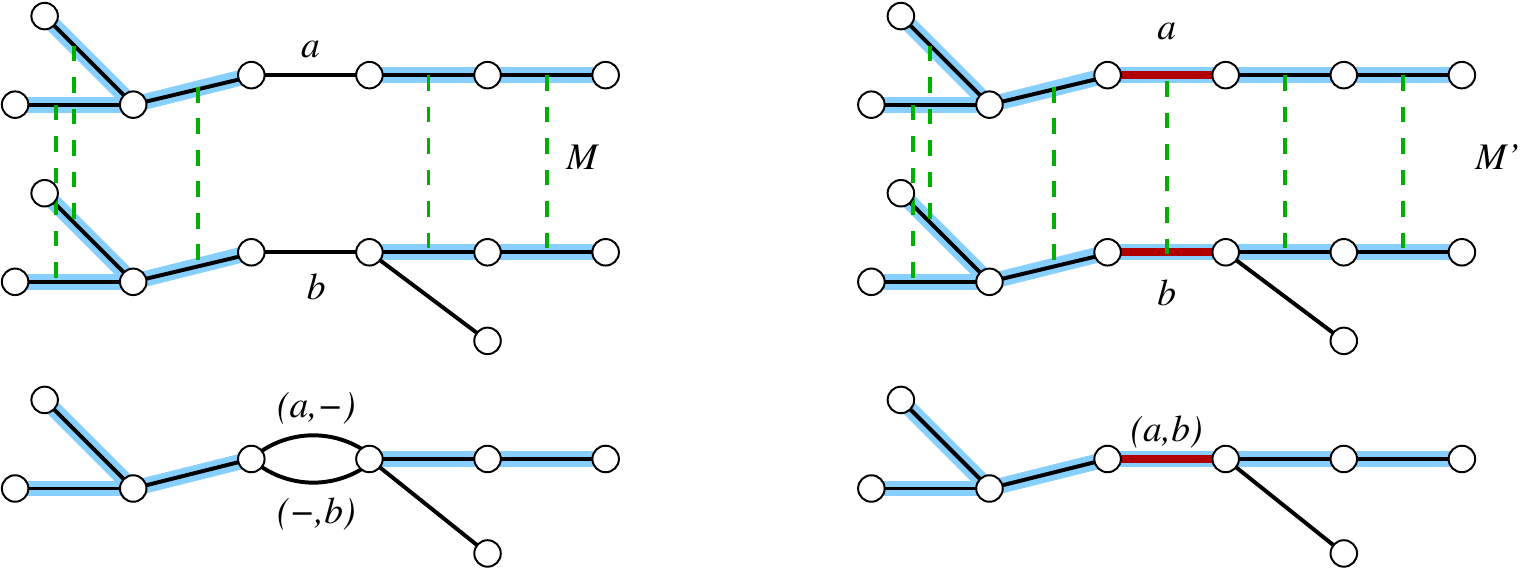}
	\end{center}
	\caption{Common sub-E-graphs and alignments. Left: The two E-graphs
		$(E_1,\mathscr{V}_1)$ and $(E_2,\mathscr{V}_2)$ are drawn as their
		corresponding graphs (top). A common sub-E-graph is indicated by the
		blue points (edges in the graph representation) and dashed ``matches''
		between them, indicating the E-graph isomorphism between the two
		subgraphs. Attempting to construct an alignment E-graph in which $a\in
		E_1$ and $b\in E_2$ have to remain unmatched, we would obtain sets
		$\Upsilon=\{(e,e),(a,-),(-,b)\}$ and
		$\Upsilon'=\{(e',e'),(a,-),(-,b)\}$, from which the sets $\{e,a\}$ and
		$\{a,e'\}$ in $(E_1,\mathscr{V}_1)$ and $\{e,b\}$ and $\{b,e'\}$ in
		$(E_2,\mathscr{V}_2)$ are recovered. However, $|\Upsilon\cap
		\Upsilon'|=2$, and thus the corresponding set system is no longer an
		E-graph. Again drawing the corresponding graph (below) this illegal
		intersection of sets can be interpreted as a double edge. Right: If the
		common sub-E-graph is extended by the match of the points $a$ and $b$,
		the corresponding sets in the alignment E-graph become
		$\Upsilon=\{(e,e),(a,b)\}$ and $\Upsilon'=\{(e',e'),(a,b)\}$, and the
		set system is again an E-graph, and thus an alignment of
		$(E_1,\mathscr{V}_1)$ and $(E_2,\mathscr{V}_2)$.  
	}
	\label{fig:implied1}
\end{figure}

In the analogous case of common induced subgraphs of two graphs this is
always the case, i.e., every common induced subgraph can be extended to an
alignment by ``gluing together'' the two input graphs at their common
induced subgraph, see \cite{Berkemer:20b}. In the case of E-graphs, the
situation is more complicated, however. The example in
Fig.~\ref{fig:implied1} shows that not every common induced sub-E-graph
gives rise to an alignment in this manner. More precisely, it is not always
possible to extend the matches-only alignment defined on a common
sub-E-graph to a valid alignment by adding only in/del columns.

To see this, consider the example in Fig.~\ref{fig:implied1}.  Since the
hypergraph representation of E-graphs is difficult to interpret, we will
instead use the corresponding graphs. It is important to keep in mind, the
points of E-graphs are drawn as edges, and thus alignment matches are
defined only between edges. Consider the common sub-E-graph indicated by
the blue points and the dashed lines denoting the E-graph isomorphism
between the subgraphs of $(E_1,\mathscr{V}_1)$ and $(E_2,\mathscr{V}_2)$.
In order to extend this common sub-E-graph to an alignment, the two points
$a$ and $b$ have to remain unmatched. This is not possible, however, since
the resulting set system is no longer an E-graph.

\begin{figure}[b]
	\begin{center}
		\includegraphics[width=0.5\columnwidth]{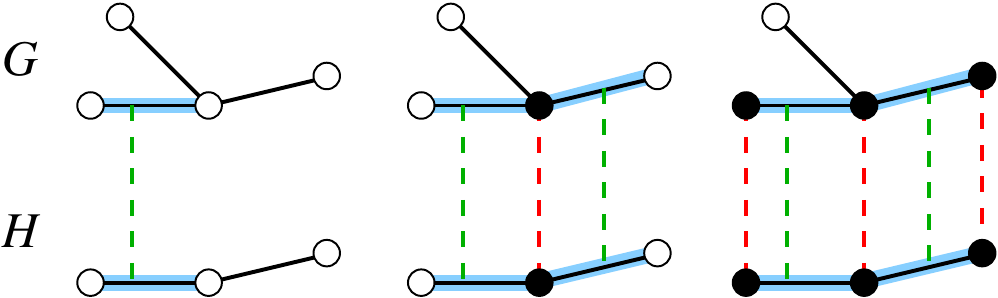}
	\end{center}
	\caption{An isolated point match (green dashed line) does not imply
		any set-matches (left). Two incident point matches, however, imply a
		match between the shared sets, shown here as the match of the shared
		vertices (red dashed line) in the corresponding graph (middle).  For
		each point match for which an incident set-match is known, the set
		match corresponding to the other end-point of graph edge is also
		implied (right).}
	\label{fig:setmatch}
\end{figure}

The situation in Fig.~\ref{fig:implied1} can be understood more concisely
by introducing \textit{set-matches} in addition to the matches of points in
E-graph.  The key observation is that whenever two points $e$ and $e'$ are
incident in an E-graph, there is a unique set $\Upsilon$ that contains both
$e$ and $e'$. In the corresponding graph, $\Upsilon$ corresponds to the
unique vertex incident to both $e$ and $e'$. Thus, any set of matches in
E-graph alignment that forms a connected component with at least two points
also uniquely defines a matching of the sets, and thus also a unique
matching of the vertices between the corresponding graphs. More precisely,
suppose $e_1'e_2'$ is a match of points $e_1'\in E_1$ and $e_2'\in E_2$ and
suppose there is $e_1\in E_1$ and $e_2\in E_2$ such that $e_1$ and $e_1'$
are incident in $(E_1,\mathscr{V}_1)$ and $e_2$ and $e_2'$ are incident in
$(E_2,\mathscr{V}_2)$. Then there are unique sets
$\Upsilon_1\in\mathscr{V}_1$ with $e_1,e_1'\in\Upsilon_1$ and
$\Upsilon_2\in\mathscr{V}_2$ with $e_2,e_2'\in\Upsilon_2$. Moreover, if
$e_1e_2$ is also a point match, then $\Upsilon_1$,$\Upsilon_2$ is
necessarily a set-match. This, in turn, implies that there are unique sets
$\Upsilon_1'\in\mathscr{V}_1$ with $e_1'\in\Upsilon_1'\ne\Upsilon_1$ and
$\Upsilon_2'\in\mathscr{V}_2$ with $e_2'\in\Upsilon_2'\ne\Upsilon_2$. These
also form a set-match. In terms of the corresponding graphs, two pairs of
incident edge-matches imply unique matches between their incident
vertices, see Fig.~\ref{fig:setmatch}.

Now consider two points $e_1\in E_1$ and $e_2\in E_2$ with
$\{e_1\}=\Upsilon_1\cap \Upsilon_1'$ and $\{e_2\}=\Upsilon_2\cap
\Upsilon_2'$ and \emph{assume} that other matches between the two E-graphs
$(E_1,\mathscr{V}_1)$ and $(E_2,\mathscr{V}_2)$ already imply the set-match
$\Upsilon_1$,$\Upsilon_2$ and $\Upsilon_1'$,$\Upsilon_2'$,
respectively. For the corresponding graphs this means that the vertices
incident with $e_1$ and $e_2$ are already matched.

In the E-graph alignment we have now two choices: (i) $e_1$ and $e_2$ form
a match column, or (ii) both $e_1$ and $e_2$ form an in/del column.  In the
latter case, the alignment contains sets $\Upsilon$ and $\Upsilon'$ from
which $\Upsilon_1$ and $\Upsilon_1'$ as well $\Upsilon_2$ and $\Upsilon_2'$
are obtained as projections, and thus both $\Upsilon$ and $\Upsilon'$ must
contain points corresponding to the in/del columns $(e_1,-)$ and
$(-,e_2)$. However, this implies $|\Upsilon\cap\Upsilon'|\ge 2$, which is
not possible in an E-graph. Thus only the first case is possible. We call
the pair of points $(e_1,e_2)$ an \emph{implied match} because this point
match is the consequence of a pair set-matches that in turn are implied by
other point matches. In terms of the corresponding graph representation
this condition can be phrased as follows: If for two edges $e_1=u_1u_1'\in
E_1$ and $e_2=u_2u_2'\in E_2$ the matching of the edges $e_1$ and $e_2$
implies (as a consequence of other edge matches) that $u_1$-$u_2$ and
$u_1'$-$u_2'$ are vertex matches, then the graph edges $e_1$ and $e_2$ must
form an  edge match, Fig.~\ref{fig:implied1}.

In order to make this reasoning precise, let $(E_1,\mathscr{V}_1)$ and
$(E_2,\mathscr{V}_2)$ be two E-graphs and let $M\subseteq E_1\times E_2$
be a set of point matches such that every point of $E_1$ and every point
of $E_2$ appears in at most one pair of $M$. We write $M_1$ and $M_2$ for
the sets of matched points in $E_1$ and $E_2$, respectively, and
$\varphi_M:M_1\to M_2$ for the bijection with $\varphi_M(e_1)=e_2$ for
$(e_1,e_2)\in M$.
\begin{definition}\label{Def:setmatch}
	The \emph{set-match relation} $\mathcal{N}(M)\subseteq
	\mathscr{V}_1\times\mathscr{V}_2$ of $M$ is the smallest relation such
	that
	\begin{description}
		\item[(N1)] $(\Upsilon_1,\Upsilon_2)\in\mathcal{N}(M)$ whenever there are
		two distinct pairs $(e_1,e_2),(e_1',e_2')\in M$ with
		$e_1,e_1'\in\Upsilon_1$ and $e_2,e_2'\in\Upsilon_2$;
		\item[(N2)] $(\Upsilon_1',\Upsilon_2')\in\mathcal{N}(M)$ whenever
		$(\Upsilon_1,\Upsilon_2)\in\mathcal{N}(M)$ and there is $(e_1,e_2)\in
		M$ with $\Upsilon_1\cap\Upsilon_1'=\{e_1\}$ and
		$\Upsilon_2\cap\Upsilon_2'=\{e_2\}$.
	\end{description}
	A pair $(e_1,e_2)\in(E_1\times E_2)\setminus M$ is an \emph{implied
		match} of $M$ if there are
	$(\Upsilon_1,\Upsilon_2),(\Upsilon_1',\Upsilon_2')\in\mathcal{N}(M)$
	with $\Upsilon_1\cap\Upsilon_1'=\{e_1\}$ and
	$\Upsilon_2\cap\Upsilon_2'=\{e_2\}$. We say that $M$ is \emph{closed} if
	it has no implied matches.
\end{definition}
Rules (N1) and (N2) are the two situations shown in the middle and right
panels of Fig.~\ref{fig:setmatch}. Note that neither rule applies to a
point match $(e_1,e_2)$ for which $e_1$ or $e_2$ is an isolated point of
its E-graph, and that no set-match is produced for a matched pair of
points that is not incident with any other matched pair.
\begin{definition}\label{Def:consistent}
	A set of point matches $M$ is \emph{consistent} if
	\begin{description}
		\item[(C1)] for all $(e_1,e_2),(e_1',e_2')\in M$, the points $e_1$ and
		$e_1'$ are incident in $(E_1,\mathscr{V}_1)$ if and only if $e_2$ and
		$e_2'$ are incident in $(E_2,\mathscr{V}_2)$, and
		\item[(C2)] $\mathcal{N}(M)$ is a partial bijection, i.e., every
		$\Upsilon_1\in\mathscr{V}_1$ and every $\Upsilon_2\in\mathscr{V}_2$
		appears in at most one pair of $\mathcal{N}(M)$.
	\end{description}
\end{definition}
Condition (C1) alone does not guarantee that $M$ defines an isomorphism
of sub-E-graphs. The three edges of a triangle $K_3$ and the three edges
of a claw $K_{1,3}$ are pairwise incident, so that every bijection between
them satisfies (C1); by rule (N1), however, the set $\{a,b,c\}$ of the claw
would have to be set-matched with each of the three sets $\{a,b\}$,
$\{b,c\}$, and $\{a,c\}$ of the triangle, violating (C2). Condition (C2) is
thus the counterpart of the explicit exclusion of $\Delta Y$ exchanges in
line-graph based approaches to the MCES problem \cite{Raymond:02}, and of
the $\Delta$-Y admissibility of cliques in the modular product of line
graphs \cite{Akbar:27}.
Conversely, (C2) does not imply (C1): two non-incident points may be
matched to two incident points without producing any set-match at all.
\begin{proposition}\label{prop:consistent}
	A set of point matches $M$ is consistent if and only if $\varphi_M$ is
	an isomorphism of the induced sub-E-graphs $(E_1,\mathscr{V}_1)[M_1]$
	and $(E_2,\mathscr{V}_2)[M_2]$. In this case,
	$(\Upsilon_1,\Upsilon_2)\in\mathcal{N}(M)$ if and only if
	$\Upsilon_1\cap M_1$ and $\Upsilon_2\cap M_2$ are sets of
	$(E_1,\mathscr{V}_1)[M_1]$ and $(E_2,\mathscr{V}_2)[M_2]$, respectively,
	with $\varphi_M(\Upsilon_1\cap M_1)=\Upsilon_2\cap M_2$.
\end{proposition}
\begin{proof}
	We first note that two points $e_1,e_1'\in M_1$ are incident in
	$(E_1,\mathscr{V}_1)$ if and only if they are incident in
	$(E_1,\mathscr{V}_1)[M_1]$: a set $\Upsilon_1\ni e_1,e_1'$ yields the set
	$\Upsilon_1\cap M_1\ni e_1,e_1'$ with at least two elements, which is
	retained by Def.~\ref{Def:sub-E-graph}, and every set of the
	sub-E-graph is of the form $\Upsilon_1\cap M_1$. Furthermore, by
	condition (ii) of Def.~\ref{Def:sub-E-graph}, a point $e_1\in M_1$ is
	not isolated in $(E_1,\mathscr{V}_1)[M_1]$ if and only if one of the two
	sets containing $e_1$ meets $M_1$ in at least two points, in which case
	both of them are retained.
	
	Suppose that $\varphi_M$ is an isomorphism. Then (C1) holds by the first
	remark. We show by induction along the derivation of $\mathcal{N}(M)$
	that every $(\Upsilon_1,\Upsilon_2)\in\mathcal{N}(M)$ satisfies
	$\varphi_M(\Upsilon_1\cap M_1)=\Upsilon_2\cap M_2$, where both
	intersections are sets of the respective sub-E-graphs. For (N1),
	$\Upsilon_1\cap M_1\ni e_1,e_1'$ is retained, and its image under
	$\varphi_M$ is a set of $(E_2,\mathscr{V}_2)[M_2]$ containing $e_2$ and
	$e_2'$; since two points of an E-graph lie in at most one common set,
	this image is $\Upsilon_2\cap M_2$. For (N2), $e_1\in\Upsilon_1\cap
	M_1$, and by the second remark both $\Upsilon_1\cap M_1$ and
	$\Upsilon_1'\cap M_1$ are retained. The image
	$\varphi_M(\Upsilon_1'\cap M_1)$ is a set of
	$(E_2,\mathscr{V}_2)[M_2]$ that contains $e_2$ and differs from
	$\varphi_M(\Upsilon_1\cap M_1)=\Upsilon_2\cap M_2$; hence it equals
	$\Upsilon_2'\cap M_2$. Since $\varphi_M$ induces a bijection between
	the sets of the two sub-E-graphs, and each set of
	$(E_1,\mathscr{V}_1)[M_1]$ is of the form $\Upsilon_1\cap M_1$ for a
	unique $\Upsilon_1\in\mathscr{V}_1$ (and analogously for
	$\mathscr{V}_2$), (C2) follows.
	
	Conversely, suppose that $M$ is consistent and let
	$\Upsilon_1\in\mathscr{V}_1$ with $|\Upsilon_1\cap M_1|\ge 2$, say
	$e_1,e_1'\in\Upsilon_1\cap M_1$. By (C1), $e_2$ and $e_2'$ lie in a
	common set $\Upsilon_2$, and $(\Upsilon_1,\Upsilon_2)\in\mathcal{N}(M)$
	by (N1). For any further $e_1''\in\Upsilon_1\cap M_1$, the point
	$e_2''$ is incident with both $e_2$ and $e_2'$. If
	$e_2''\notin\Upsilon_2$, then $e_2$ and $e_2''$ lie in a common set
	$\Upsilon_2''\ne\Upsilon_2$, and (N1) yields
	$(\Upsilon_1,\Upsilon_2'')\in\mathcal{N}(M)$, contradicting (C2). Hence
	$\varphi_M(\Upsilon_1\cap M_1)\subseteq\Upsilon_2\cap M_2$, and the same
	argument applied to $\varphi_M^{-1}$ shows equality. Now let
	$\Upsilon_1'\cap M_1=\{e_1\}$ be a set of $(E_1,\mathscr{V}_1)[M_1]$
	retained by condition (ii) of Def.~\ref{Def:sub-E-graph}, i.e.,
	$\Upsilon_1\cap\Upsilon_1'=\{e_1\}$ with $|\Upsilon_1\cap M_1|\ge 2$. We
	have just seen that $\varphi_M(\Upsilon_1\cap M_1)=\Upsilon_2\cap M_2$,
	so that $\Upsilon_2'\cap M_2\ni e_2$ is retained in
	$(E_2,\mathscr{V}_2)[M_2]$, where $\Upsilon_2\cap\Upsilon_2'=\{e_2\}$.
	If $\Upsilon_2'\cap M_2$ contained a second point $f_2=\varphi_M(f_1)$,
	then $f_1$ would be incident with $e_1$ by (C1), hence
	$f_1\in\Upsilon_1\cap M_1$ and thus $f_2\in\Upsilon_2\cap\Upsilon_2'=
	\{e_2\}$, a contradiction. Thus $\varphi_M(\{e_1\})=\Upsilon_2'\cap
	M_2$, and again the symmetric argument applies to $\varphi_M^{-1}$.
	Since every set of $(E_1,\mathscr{V}_1)[M_1]$ is of one of the two
	types considered, $\varphi_M$ maps the sets of
	$(E_1,\mathscr{V}_1)[M_1]$ bijectively onto the sets of
	$(E_2,\mathscr{V}_2)[M_2]$, i.e., it is an isomorphism. The
	characterization of $\mathcal{N}(M)$ follows from the first part of the
	proof together with the observation that every retained set is reached
	by (N1) or (N2).
\end{proof}
In particular, the consistent sets of point matches are exactly the
common induced sub-E-graphs of $(E_1,\mathscr{V}_1)$ and
$(E_2,\mathscr{V}_2)$, and $\mathcal{N}(M)$ is the matching of sets,
i.e., of vertices of the corresponding graphs, that is determined by
$M$. The matched points that are isolated in the common sub-E-graph are
those for which $\mathcal{N}(M)$ contains no information; we return to
them in Sect.~\ref{sec:pairwise}. A subset of a consistent set of point
matches is again consistent, since (C1) is inherited by subsets and
$\mathcal{N}(M')\subseteq\mathcal{N}(M)$ for $M'\subseteq M$.

The argument given above for the example of Fig.~\ref{fig:implied1} now
generalizes as follows. Since multiple alignments are composed from
pairwise alignments of E-graphs that may themselves be alignments, it
suffices to consider the pairwise case.
\begin{lemma}\label{lem:implied}
	Let $M$ be a consistent set of point matches between the E-graphs
	$(E_1,\mathscr{V}_1)$ and $(E_2,\mathscr{V}_2)$ such that the common
	sub-E-graph $(E_1,\mathscr{V}_1)[M_1]$ has no isolated points. Then
	there is an alignment of $(E_1,\mathscr{V}_1)$ and $(E_2,\mathscr{V}_2)$
	whose set of match columns is $M$ if and only if $M$ is closed under
	implied matches.
\end{lemma}
\begin{proof}
	Suppose $(E,\mathscr{V})$ is an alignment with monomorphisms
	$\mu_1,\mu_2$ whose set of match columns is $M$, i.e.,
	$\mu_1(e_1)=\mu_2(e_2)$ if and only if $(e_1,e_2)\in M$. We first
	observe that $(\Upsilon_1,\Upsilon_2)\in\mathcal{N}(M)$ implies that
	there is a set $\Upsilon\in\mathscr{V}$ with
	$\mu_1^{-1}(\Upsilon)=\Upsilon_1$ and $\mu_2^{-1}(\Upsilon)=\Upsilon_2$.
	For (N1), the two match columns $\mu_1(e_1)=\mu_2(e_2)$ and
	$\mu_1(e_1')=\mu_2(e_2')$ lie in a unique common set
	$\Upsilon\in\mathscr{V}$; since $(E,\mathscr{V})[\mu_1(E_1)]\simeq
	(E_1,\mathscr{V}_1)$, its preimage $\mu_1^{-1}(\Upsilon)$ is the unique
	set of $\mathscr{V}_1$ containing $e_1$ and $e_1'$, i.e., $\Upsilon_1$,
	and analogously $\mu_2^{-1}(\Upsilon)=\Upsilon_2$. For (N2), the match
	column $x=\mu_1(e_1)=\mu_2(e_2)$ is contained in exactly two sets of
	$\mathscr{V}$, one of which is the set $\Upsilon$ corresponding to
	$(\Upsilon_1,\Upsilon_2)$; the other one therefore corresponds to
	$(\Upsilon_1',\Upsilon_2')$. Now let $(e_1,e_2)$ be an implied match,
	with $(\Upsilon_1,\Upsilon_2),(\Upsilon_1',\Upsilon_2')\in\mathcal{N}(M)$
	corresponding to $\Upsilon,\Upsilon'\in\mathscr{V}$. Then
	$\mu_1(e_1),\mu_2(e_2)\in\Upsilon\cap\Upsilon'$, and
	$|\Upsilon\cap\Upsilon'|\le 1$ forces $\mu_1(e_1)=\mu_2(e_2)$, i.e.,
	$(e_1,e_2)\in M$, contradicting the definition of an implied
	match. Hence $M$ is closed.
	
	Conversely, suppose that $M$ is closed. By Prop.~\ref{prop:consistent},
	$\mathcal{N}(M)$ is a partial bijection and, since the common
	sub-E-graph has no isolated points, both sets containing a matched point
	$e_1\in M_1$ are set-matched to the two sets containing
	$\varphi_M(e_1)$. We construct $E$ as the disjoint union of $M$,
	$E_1\setminus M_1$, and $E_2\setminus M_2$, with $\mu_1(e_1)=(e_1,e_2)$
	for $(e_1,e_2)\in M$ and $\mu_1(e_1)=e_1$ otherwise, and analogously
	for $\mu_2$. For each $(\Upsilon_1,\Upsilon_2)\in\mathcal{N}(M)$ we
	define $\Upsilon\coloneqq\mu_1(\Upsilon_1)\cup\mu_2(\Upsilon_2)$, and
	for each set $\Upsilon_\alpha\in\mathscr{V}_\alpha$ that does not appear
	in $\mathcal{N}(M)$ we define $\Upsilon\coloneqq\mu_\alpha(\Upsilon_\alpha)$,
	$\alpha=1,2$; let $\mathscr{V}$ be the collection of these sets. Every
	point of $E$ is contained in exactly two or in none of the sets of
	$\mathscr{V}$: for a match $(e_1,e_2)$ this follows from the fact that
	the two sets containing $e_1$ are set-matched to the two sets
	containing $e_2$; for an unmatched point it is inherited from its
	E-graph. Let $\Upsilon,\Upsilon'\in\mathscr{V}$ be distinct. If both
	correspond to set-matches $(\Upsilon_1,\Upsilon_2)$ and
	$(\Upsilon_1',\Upsilon_2')$, then $\Upsilon\cap\Upsilon'$ consists of the
	points $\mu_1(\Upsilon_1\cap\Upsilon_1')$ and
	$\mu_2(\Upsilon_2\cap\Upsilon_2')$, where $|\Upsilon_1\cap\Upsilon_1'|\le
	1$ and $|\Upsilon_2\cap\Upsilon_2'|\le 1$. If
	$\Upsilon_1\cap\Upsilon_1'=\{e_1\}$ and $\Upsilon_2\cap\Upsilon_2'=\{e_2\}$
	with $(e_1,e_2)\notin M$, then $(e_1,e_2)$ would be an implied match,
	which is excluded; hence either one of the two intersections is empty,
	or $(e_1,e_2)\in M$ and $\mu_1(e_1)=\mu_2(e_2)$, so that
	$|\Upsilon\cap\Upsilon'|\le 1$ in all cases. If at least one of
	$\Upsilon,\Upsilon'$ is the image of a single set $\Upsilon_\alpha$, then
	$|\Upsilon\cap\Upsilon'|\le|\Upsilon_\alpha\cap\mu_\alpha^{-1}(\Upsilon')|
	\le 1$. Thus $(E,\mathscr{V})$ is an E-graph. By construction,
	$\mu_\alpha^{-1}(\Upsilon)$ is either empty or a set of
	$\mathscr{V}_\alpha$, every set of $\mathscr{V}_\alpha$ arises in this way
	from exactly one $\Upsilon\in\mathscr{V}$, and a set
	$\Upsilon_\alpha=\mu_\alpha^{-1}(\Upsilon)$ is retained in the
	sub-E-graph $(E,\mathscr{V})[\mu_\alpha(E_\alpha)]$ precisely when
	$|\Upsilon_\alpha|\ge 2$ or $\Upsilon_\alpha=\{e\}$ with $e$ non-isolated
	in $(E_\alpha,\mathscr{V}_\alpha)$, i.e., always. Hence
	$(E,\mathscr{V})[\mu_\alpha(E_\alpha)]\simeq(E_\alpha,\mathscr{V}_\alpha)$
	for $\alpha=1,2$, and $(E,\mathscr{V})$ is an alignment with set of match
	columns $M$.
\end{proof}
Since E-graph alignments are again E-graphs, Lemma~\ref{lem:implied}
applies verbatim to the alignment of two E-graph alignments and thus
covers the progressive construction of multiple alignments in
Sect.~\ref{sec:progressive}. If the common sub-E-graph has isolated
points, the construction in the proof requires an additional choice for
each isolated match; this is discussed in Sect.~\ref{sec:pairwise}.

\subsection{Ambiguous Sets}\label{sec:ambiguous}

\begin{figure}
	\centering
	\includegraphics[width=0.55\columnwidth]{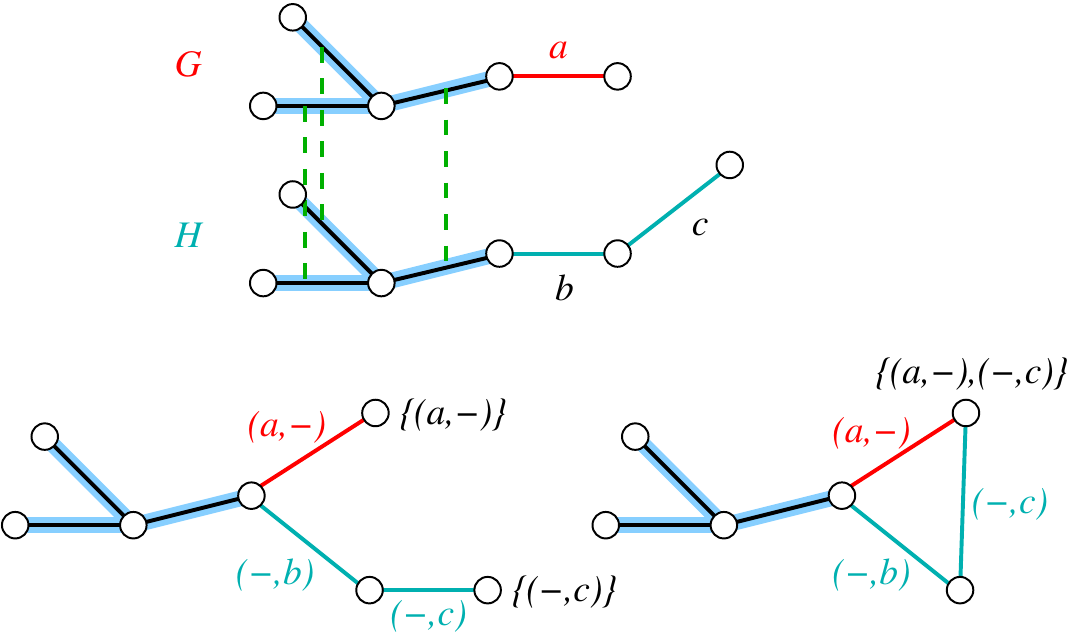}
	
	\caption{Ambiguous sets in an E-graph alignment. The common sub-E-graph
		of the two E-graphs $G$ and $H$ (matches shown by dashed vertical
		lines) leaves the points $a$ in $G$ and the points $b$ and $c$ in $H$
		unmatched. The two E-graph alignments (below) can be constructed
		depending on whether $a$ and $c$ are considered incident in the
		alignment (right, there is the set $\{(a,-),(-,c)\}$) or not (left,
		$\{(a,-)\}$ and $\{(-,c)\}$ appear as distinct sets).
	}
	\label{fig:ambiguous}
\end{figure}

In the case of vertex-centric graph alignments the same common induced
subgraph can give rise to multiple graph alignments. This is the case
whenever two alignment columns have no rows in common. In this case it does
not matter whether the alignment contains an edge between them, since in
each row at least one of the two vertices is deleted, and thus such an edge
in the alignment graph never survives projection to any of the rows. An
analogous issue appears for E-graph alignments in the form of
\emph{ambiguous sets} whenever there are two points in the alignment that
never appear in the same row. Fig.~\ref{fig:ambiguous} shows such an
example for pairwise E-graph alignments.

In order to formalize this idea, we consider an alignment E-graph
$(E,\mathscr{V})$ with rows $(E_{\alpha},\mathscr{V}_{\alpha})$, $\alpha\in
S$. For $\Upsilon\in\mathscr{V}$ we write
\begin{equation} 
	\supp(\Upsilon)\coloneqq \{\alpha\in S \ | \ \exists e\in\Upsilon:
	\mu_{\alpha}^{-1}(e)\ne\emptyset\}
\end{equation}
for the ``support'' of $\Upsilon$, i.e., the set of alignment rows that are
represented in at least one point $e\in\Upsilon$. We say that two sets
$\Upsilon_1,\Upsilon_2\in\mathscr{V}$ are \emph{mergeable} if
$\supp(\Upsilon_1)\cap\supp(\Upsilon_2)=\emptyset$ and
  $|\Lambda\cap(\Upsilon_1\cup\Upsilon_2)|\le 1$ for all
  $\Lambda\in\mathscr{V}\setminus\{\Upsilon_1,\Upsilon_2\}$, i.e., if no
  third set meets both $\Upsilon_1$ and $\Upsilon_2$. Note that
$\Upsilon_1\cap\Upsilon_2=\emptyset$ whenever the supports are
  disjoint, but the converse is of course not true. In terms of the
  corresponding graphs, two vertices are mergeable if they do not share a
  row and have no common neighbor; identifying two vertices with a common
  neighbor would create a double edge as in Fig.~\ref{fig:implied1}. Set
$\mathscr{V}'\coloneqq \mathscr{V}\setminus \{\Upsilon_1,\Upsilon_2\}
\cup\{\Upsilon_1\cup\Upsilon_2\}$, i.e., to obtain $\mathscr{V}'$ we
replace $\Upsilon_1$ and $\Upsilon_2$ by their union. Then
  $(E,\mathscr{V}')$ is again an E-graph: every point of
  $\Upsilon_1\cup\Upsilon_2$ is still contained in exactly two sets because
  $\Upsilon_1\cap\Upsilon_2=\emptyset$, and
  $|\Lambda\cap(\Upsilon_1\cup\Upsilon_2)|\le 1$ holds by assumption.
  Moreover, if $(E,\mathscr{V})$ is an alignment of the E-graphs
$(E_\alpha,\mathscr{V}_{\alpha})$ then this remains true for
$(E,\mathscr{V}')$ since for each $\alpha$, we have
$\mu_{\alpha}^{-1}(\Upsilon_1)=\emptyset$ or
$\mu_{\alpha}^{-1}(\Upsilon_2)=\emptyset$, so that the restriction of
  $(E,\mathscr{V}')$ to every row coincides with that of
  $(E,\mathscr{V})$.  Conversely, suppose that a set
$\Upsilon'\in\mathscr{V}$ can be partitioned into two non-empty subsets
$\Upsilon_1'$ and $\Upsilon_2'$ with disjoint supports. Replace
  $\Upsilon'$ by these two sets to obtain $\mathscr{W}$, and suppress every
  singleton $\{e\}\in\mathscr{W}$ for which no set of $\mathscr{W}$ of
  cardinality at least two contains $e$.  The resulting set system is an
  alternative E-graph alignment of the same rows; suppression removes the
  endpoints of isolated edges created by the split. If both split sets
  survive, they are mergeable. We call $\Upsilon'$ \emph{splittable}.

\subsection{Monotone Scoring Functions}\label{sec:scoring}

Let us first consider a pairwise E-graph alignment $(E,\mathscr{V})$ and
denote by $M\subseteq E$ the set of matches, i.e., $e\in M$ if and only if
$\mu_{1}^{-1}(e)\ne \emptyset$ and $\mu_{2}^{-1}(e)\ne \emptyset$. Since
$M$ implicitly determines the set of insertions and deletions, $\{e\in
E\ |\ \mu_{1}^{-1}(e)=\emptyset \text{ or } \mu_{2}^{-1}(e)=\emptyset\}$,
a scoring function for pairwise alignments can be written as a function
$f(M)$ of the set of matches alone. The most common choice is additive,
\begin{equation}\label{eq:pairscore}
  f(M)=\sum_{e\in M}\sigma\big(\lambda_1(\mu_1^{-1}(e)),
  \lambda_2(\mu_2^{-1}(e))\big) + \gamma\,\big(|E_1|+|E_2|-2|M|\big)\,,
\end{equation}
where $\lambda_\alpha:E_\alpha\to\mathscr{L}$ assigns labels to the points
of the input E-graphs, $\sigma:\mathscr{L}\times\mathscr{L}\to\mathbb{R}$
scores a (mis)match of labels, and $\gamma\le 0$ is the score of an in/del
column; note that $|E_1|+|E_2|-2|M|$ is the number of in/del columns. A
scoring function is strictly monotone if $M'\subsetneq M$ implies
$f(M')<f(M)$. The additive score \eqref{eq:pairscore} is strictly
  monotone if and only if $\sigma(x,y)>2\gamma$ for all
  $x,y\in\mathscr{L}$, since adding a match replaces two in/del columns by
  one match column. Label combinations that are forbidden altogether, as in
  Sect.~\ref{sec:constraints}, are conveniently modeled by
  $\sigma(x,y)=-\infty$ or, equivalently, by excluding the corresponding
  pairs from the candidate matches; strict monotonicity then refers to the
  allowed matches only. A set of matches is maximal if it cannot be
extended by an allowed match. Similarly, $M$ is maximal w.r.t.\ score
if adding (or removing) a point does not increase the score.

For strictly monotone scoring functions it follows immediately that a
set of matches $M$ is score maximal only if it is maximal w.r.t.\
cardinality. In particular, the maximum score alignment can be found by
identifying, among all cardinality maximal sets of matches, one that
maximizes the score, see also \cite{Bunke:97,Brun:12}.

For alignments of multiple E-graphs the so-called sum of pairs score is a
natural choice \cite{Carillo:88}. That is, for every pair
$\alpha,\beta\in S$, $\alpha\ne\beta$, one considers the set of matches
$M_{\alpha\beta}\coloneqq\{e\in E\ |\ \mu_{\alpha}^{-1}(e)\ne\emptyset
  \text{ and } \mu_{\beta}^{-1}(e)\ne\emptyset\}$ between rows $\alpha$
and $\beta$. The score of the alignment is given by
\begin{equation}\label{eq:SP}
    \mathrm{SP}(E,\mathscr{V})=\sum_{\{\alpha,\beta\}\subseteq S,\
      \alpha\ne\beta} f(M_{\alpha\beta})\,,
  \end{equation}
  where $f(M_{\alpha\beta})$ is evaluated according to
  Eq.~\eqref{eq:pairscore} for the pairwise alignment of rows $\alpha$ and
  $\beta$, i.e., for the restriction of $(E,\mathscr{V})$ to the rows
  $\{\alpha,\beta\}$ from which gap-only columns have been removed.
  Equivalently, a column $e\in E$ contributes
  $\sigma(\Lambda_\alpha(e),\Lambda_\beta(e))$ for every pair of rows in
  which it is present, $\gamma$ for every pair of rows of which exactly one
  is present, and nothing for pairs of rows in which it is absent, where
  $\Lambda_\alpha(e)$ denotes the label of $\mu_\alpha^{-1}(e)$. The
  sum-of-pairs score may be normalized by the number of row pairs. Scores
  that depend on set-matches, i.e., on vertex labels, are discussed in
  Sect.~\ref{sec:constraints}.

\section{Algorithms}

\subsection{VF2 Algorithms}

In its basic form, the VF2 algorithm \cite{Cordella:04} addresses the
subgraph isomorphism problem: Given a pattern graph $H$, does a target $G$
contain a subgraph that is isomorphic to $H$? To this end, it
systematically constructs mappings $\phi:V(H)\to V(G)$ such that for every
edge $uv\in E(H)$, $\phi(u)\phi(v)\in E(G)$. The basic idea is to
stepwisely extend a partial mapping $\phi:W\to V(G)$, where $W\subseteq
V(H)$. The same idea pertains to induced subgraphs by requiring instead
that $H[W]\simeq G[\phi(W)]$, which amounts to keeping track of non-edges
as well. VF2 utilizes a recursive exploration and backtracking strategy
that guarantees the comprehensive enumeration of solutions and thus exact
solutions at the expense of exponential worst-case running times.  Each
partial map $\phi:V(H)\to V(G)$ is represented by a list $\mathcal{M}$ of
pairs of vertices $(u,\phi(u))$.
\begin{itemize}
	\item[(i)] VF2 starts from an  empty mapping $\mathcal{M}=\emptyset$.
	\item[(ii)] For each $u\in V(H)$ that has not yet been mapped, a set
	$P(u)\subseteq V(G)$ of \emph{candidate matches} is generated based on
	structural and attribute constraints.
	\item[(iii)] For each $v\in P(u)$, VF2 checks if the current mapping
	$\mathcal{M}$ can be \emph{extended} by $(u,v)$. This step in particular
	checks whether edges and possibly non-edges between $u$ and its
	neighbors among the already mapped vertices of $H$ are preserved in
	their respective images in $G$.
	\item[(iv)] A \emph{consistency check} determines the validity of the
	extension $\mathcal{M}\cup\{(u,v)\}$.
\end{itemize}
If the partial mapping is consistent, VF2 recursively explores further
matches by extending $\phi$ with more pairs of vertices.  This proceeds
until either $V(H)$ is mapped completely or no further extension is
possible. VF2 then backtracks and attempts a different extension of the
previous partial mapping. This amounts to a search tree whose vertices are
the partial mappings $\mathcal{M}$, eventually producing all solutions as
leaves of this tree. In its modern implementations the VF2 algorithm
incorporates various heuristics throughout its execution, such as degree
ordering, candidate pair generation prioritization, look-ahead strategies,
and pruning techniques, which guide the exploration process and enhance
efficiency \cite{Carletti:15,Juettner:18}. Since leaves in the search tree
also correspond to (maximal) incomplete mappings, VF2 and its variants can
also be used to solve maximum common subgraph problems. This is used
e.g.\ for the connected maximum common subgraph problem in the Small
Molecule Subgraph Detector (SMSD) toolkit \cite{Rahman:09} and similar
applications \cite{Cao:08,Schmidt:21}. Below, we shall see that the core
strategies of VF2 remain applicable to set systems beyond graphs.

\begin{algorithm}
  \caption{Procedure $\textrm{Match}(\mathcal{M},\mathcal{P})$}
  \label{alg:BKstyle1}
  \begin{algorithmic}[1]
      \IF{$\textrm{compatible}(\mathcal{M},p)$ is false for all $p\in\mathcal{P}$}
         \STATE \textbf{report} $\mathcal{M}$ and \textbf{return}
      \ENDIF
      \FORALL{$p\in \mathcal{P}$ in a fixed order}
         \STATE $\mathcal{P}\leftarrow \mathcal{P}\setminus \{p\}$
         \IF{$\textrm{compatible}(\mathcal{M},p)$}
            \STATE $\text{Match}\big(\mathcal{M}\cup\{p\},\
            \textrm{update}(\mathcal{P},\mathcal{M}\cup\{p\})\big)$
         \ENDIF
      \ENDFOR
  \end{algorithmic}
\end{algorithm}

\subsection{Maximal Pairwise Common Sub-E-graphs}\label{sec:maxcommon}

Let us consider the task of computing an alignment of two E-graphs
$(E_1,\mathscr{V}_1)$ and $(E_2,\mathscr{V}_2)$ using a strictly monotone
scoring function. As noted in the section on scoring functions, in
this case it suffices to enumerate all the maximal common sub-E-graphs, and
to select the highest-scoring one. Moreover, maximal common sub-E-graphs
cannot have further implied matches, since these correspond to
additional matches and hence would contradict
maximality. Indeed, if $(e_1,e_2)$ is an implied match of a consistent
  set $\mathcal{M}$, then $\mathcal{M}\cup\{(e_1,e_2)\}$ is again
  consistent, with
  $\mathcal{N}(\mathcal{M}\cup\{(e_1,e_2)\})=\mathcal{N}(\mathcal{M})$: by
  Prop.~\ref{prop:consistent}, the matched points incident with $e_1$ are
  exactly those in the two set-matched sets containing $e_1$, and their
  partners are exactly the matched points in the two corresponding sets
  containing $e_2$, so that (C1) holds for the new pair, and the new pair
  does not generate set-matches other than the two that imply it.
In the following we adapt the general recursive backtracking procedure
underlying VF2, see Alg.~\ref{alg:BKstyle1}, to solve the pairwise E-graph
alignment problem exactly.

At the root of the search tree there are no matches, i.e.\
$\mathcal{M}=\emptyset$ and the set of candidate edge matches is
$\mathcal{P}=E_1\times E_2$. The leaves of the search tree have been
reached whenever there is no further candidate that can extend
$\mathcal{M}$.  Since $p$ is removed from $\mathcal{P}$ before the
  recursive call, every set of matches is generated at most once, namely in
  the order in which its elements appear in the fixed ordering of the
  candidates. As a consequence, a set $\mathcal{M}$ reported at a leaf is
  maximal with respect to the candidates remaining in $\mathcal{P}$, but
  not necessarily with respect to candidates removed at earlier branching
  points; such non-maximal leaves are subsets of sets reported at other
  leaves and are never score-optimal for strictly monotone scoring
  functions. If only maximal sets are to be reported, the usual device of
  Bron--Kerbosch-type enumeration \cite{Bron:73} applies: the set of
  candidates removed at earlier branching points is passed along, and
  $\mathcal{M}$ is reported only if none of them is compatible with
  $\mathcal{M}$. We note that the enumeration of maximal consistent sets of
  matches could alternatively be organized as a clique enumeration in a
  suitably defined product of the two E-graphs, in the same way as the
  modified Bron--Kerbosch procedure of \cite{Akbar:27} enumerates maximal
  $\Delta$-Y admissible cliques in the modular product of line graphs; we
  have chosen the VF2-like formulation here because the set-matches are
  naturally maintained incrementally along the search tree. A solution
with maximum score is then selected.

Of course, the compatibility test and the rules for updating the
candidate list $\mathcal{P}$ need to be specified.  Let
$\mathcal{M}\subseteq E_1\times E_2$ be a  consistent set of point matches
and let $p\coloneqq(e_1,e_2)\in\mathcal{P}$ be a candidate match. Then
$\mathcal{M}\cup\{p\}$ is compatible if it is consistent in the sense of
Def.~\ref{Def:consistent}, i.e., if
\begin{itemize}
\item[(1)] for all $q=(e_1',e_2')\in\mathcal{M}$, $e_1$ and $e_1'$ are
  incident in $E_1$ if and only if $e_2$ and $e_2'$ are incident in
  $E_2$, and
\item[(2)] the set-match relation $\mathcal{N}(\mathcal{M}\cup\{p\})$
    is a partial bijection.
\end{itemize}
As shown in Sect.~\ref{sec:egraphalign}, the incidence test (1) alone
  would accept, e.g., any bijection between the edges of a triangle and the
  edges of a claw; condition (2) excludes such matchings.  The update of
$\mathcal{P}$ then at least removes all candidates containing $e_1$ or
$e_2$. Since the candidates are tested one at a time and subsets of
  consistent sets are consistent, the nodes of the search tree are exactly
  the consistent sets of point matches (each generated in a single
  insertion order); in particular, every maximal consistent set, i.e., by
  Prop.~\ref{prop:consistent}, every maximal common induced sub-E-graph, is
  visited.

\subsection{Pairwise E-graph Alignments}\label{sec:pairwise}

Let $(E_1,\mathscr{V}_1)$ and $(E_2,\mathscr{V}_2)$ be two E-graphs and let
$\mathcal{M}\subseteq E_1\times E_2$ be a cardinality-maximal set of
point-matches. Then $(E,\mathscr{V})$ is an alignment for $\mathcal{M}$ if
$\mu_1(e_1)=\mu_2(e_2)\in E$ if and only if $(e_1,e_2)\in\mathcal{M}$. It
will be convenient to keep track of matches not only between points, i.e.,
the $\mathcal{M}$, but also between the sets. Denote by
$\mathcal{N}=\mathcal{N}(\mathcal{M})$ the set of set-matches, Def.~\ref{Def:setmatch}. We observe:
\begin{itemize}
	\item[(i)] If $e_1,e_1'\in\Upsilon_1$, $e_2,e_2'\in\Upsilon_2$,
	$(e_1,e_2)\in\mathcal{M}$, and $(e_1',e_2')\in\mathcal{M}$, then
	$(\Upsilon_1,\Upsilon_2)\in \mathcal{N}$.
	\item[(ii)] If $e_1\in\Upsilon_1\cap\Upsilon_1'$,
	$e_2\in\Upsilon_2\cap\Upsilon_2'$, $(e_1,e_2)\in\mathcal{M}$, and
	$(\Upsilon_1,\Upsilon_2)\in\mathcal{N}$, then
	$(\Upsilon_1',\Upsilon_2')\in\mathcal{N}$.
	\item[(iii)] If $e_1\in\Upsilon_1\cap\Upsilon_1'$,
	$e_2\in\Upsilon_2\cap\Upsilon_2'$,
	$(e_1,e_2)\notin\mathcal{M}$, and
	$(\Upsilon_1,\Upsilon_2)\in\mathcal{N}$, then
	$(\Upsilon_1',\Upsilon_2')\notin\mathcal{N}$.
\end{itemize}
Observations (i) and (ii) are the rules (N1) and (N2) of
  Def.~\ref{Def:setmatch}, and (iii) follows from closedness under implied
  matches, which holds here by maximality (Sect.~\ref{sec:maxcommon}).  In
particular, therefore, a match of isolated points in $(E_1,\mathscr{V}_1)$
and $(E_2,\mathscr{V}_2)$ does not give rise to set-matches because the
isolated points are not associated with sets in the first place. Therefore,
they also do not contribute to implied point matches.  In summary, no
set-matches are produced for sets without incident point matches. Since
set-matches appear in $ \mathcal{N}$ if there are at least two incident
point-matches with points present in both E-graph $(E_1,\mathscr{V}_1)$ and
$(E_2,\mathscr{V}_2)$, none of the set-matches in $ \mathcal{N}$ is a
splittable set in the alignment E-graph.  Moreover, the set-matches in
$\mathcal{N}$ then may give rise to implied point-matches:
\begin{itemize}
\item[(iv)] If $e_1\in\Upsilon_1\cap\Upsilon_1'$,
  $e_2\in\Upsilon_2\cap\Upsilon_2'$,
  $(\Upsilon_1,\Upsilon_2)\in\mathcal{N}$, and
  $(\Upsilon_1',\Upsilon_2')\in\mathcal{N}$, then
  $(e_1,e_2)\in\mathcal{M}$.
\end{itemize}
These implied point-matches clearly cannot produce additional set-matches,
since they correspond to graph edges with both incident graph vertices
already fixed in set-matches. Thus a given set of point matches
$\mathcal{M}$ can be completed by first determining its set-matches and
then adding the implied point matches. After completion, therefore,
$\mathcal{N}$ contains no splittable set-matches.

\begin{figure*}
  \begin{center}
    \includegraphics[width=0.75\textwidth]{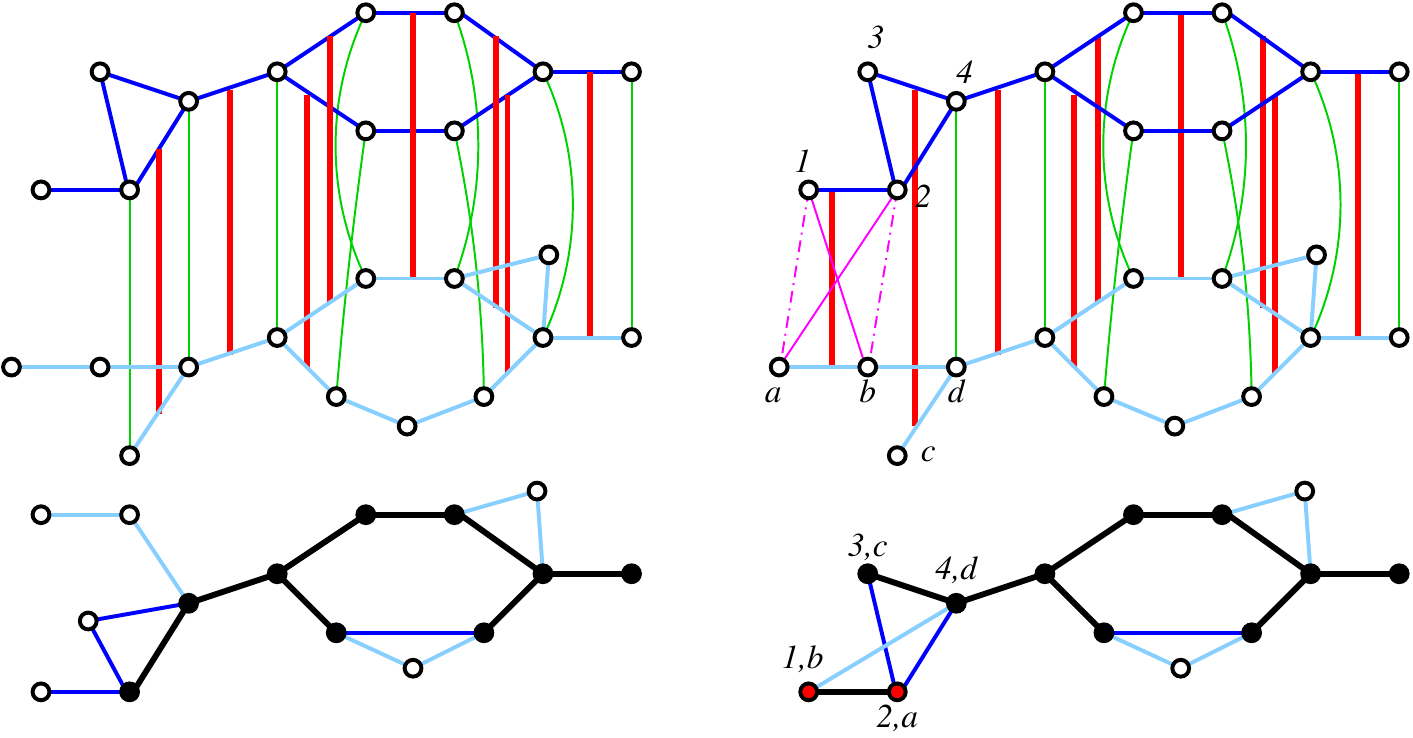}
    \newline
    \hspace*{-1.5cm}(a)\hspace*{0.375\textwidth}(b)
  \end{center}
  \caption{Examples of a pairwise alignment of two E-graphs. Points, i.e.,
    graph edges, are shown in light and dark blue, and matches of points
    are shown by vertical red lines. Set-matches (i.e., matches between the
    corresponding graph vertices) are shown in green. The sets of point
    matches differ only to the very left.  Below, the graphs
    corresponding to the alignment E-graphs are shown with
    point matches as black edges, while unmatched points are
    shown as dark or light blue edges depending on
    whether they are projected into the first or second input
    E-graph. Set matches are shown as filled vertices, unmatched sets
    are represented by open circles.  There are no isolated point
    matches in (a) and hence all set matches are implied. An isolated point
    match of the graph edges $\{1,2\}$ and $\{a,b\}$ is present in panel
    (b). \textit{A priori} there are two possible set matches, indicated in
    magenta.  Only the orientation with the set matches $1,b$ and $2,a$,
    however, leads to a valid alignment, shown below as its corresponding
    graph. The alternative choice, i.e., the set matches $1,a$ and $2,b$,
    together with the set match $4,d$ (which is implied by the remaining
    point-matches), would imply the additional point match $\{4,d;2,b\}$
    (see also Fig.~\ref{fig:implied1}). This point match is not
    present in our example. Thus only the set matches $1,b$ and $2,a$ are
    possible. The E-graph of the alignment therefore translates into the
    unique graph shown. We note that this situation may also appear in
    score-optimal E-graph alignments. For instance, the point match
    $\{4,d;2,b\}$ might be excluded by a scoring function based on the
    combinations of labels associated with points $\{4,2\}$ and
    $\{b,d\}$; without such a constraint, the set of point matches in
      (b) would not be maximal, see Sect.~\ref{sec:pairwise}.  }
  \label{fig:Egral}
\end{figure*}

A choice remains to be made if the common sub-E-graph induced by
$\mathcal{M}$ contains isolated points. Let $I\subseteq\mathcal{M}$ be the
set of \emph{isolated matches}, i.e., of pairs $(e_1,e_2)\in\mathcal{M}$
such that $e_1$ is isolated in $(E_1,\mathscr{V}_1)[M_1]$ and,
equivalently, $e_2$ is isolated in $(E_2,\mathscr{V}_2)[M_2]$. If $e_1$ or
$e_2$ is an isolated point of its input E-graph, nothing needs to be
decided: the sets containing the other point, if any, simply remain
unmatched, and their projections onto the other row are not retained.
Otherwise, $e_1\in\Upsilon_1\cap\Upsilon_1'$ and
$e_2\in\Upsilon_2\cap\Upsilon_2'$, and an \emph{orientation} of
$(e_1,e_2)$ is the choice of one of the two possible pairs of
set-matches, $\{(\Upsilon_1,\Upsilon_2),(\Upsilon_1',\Upsilon_2')\}$ or
$\{(\Upsilon_1,\Upsilon_2'),(\Upsilon_1',\Upsilon_2)\}$. In the
corresponding graphs this is the choice of which endpoints of the two
matched edges are identified, cf.\ Fig.~\ref{fig:Egral}(b). Denote by
$\mathcal{N}^{\star}$ the extension of $\mathcal{N}(\mathcal{M})$ by the
set-matches of a chosen orientation for every $(e_1,e_2)\in I$ with
non-isolated $e_1$ and $e_2$. Since $e_1$ is isolated in the common
sub-E-graph, neither $\Upsilon_1$ nor $\Upsilon_1'$ (and, likewise,
neither $\Upsilon_2$ nor $\Upsilon_2'$) appears in
$\mathcal{N}(\mathcal{M})$, and since distinct isolated matches are
pairwise non-incident by (C1), their sets are pairwise distinct. Hence
$\mathcal{N}^{\star}$ is a partial bijection for every choice of
orientations. We call a choice of orientations \emph{admissible} if, in
addition, every point match implied by $\mathcal{N}^{\star}$ (in the
sense of Def.~\ref{Def:setmatch}, with $\mathcal{N}^{\star}$ in place of
$\mathcal{N}(\mathcal{M})$) belongs to $\mathcal{M}$. The construction in
the proof of Lemma~\ref{lem:implied}, applied with $\mathcal{N}^{\star}$
instead of $\mathcal{N}(\mathcal{M})$, then yields an alignment with set
of matches $\mathcal{M}$, 
and every alignment with set of matches $\mathcal{M}$ and no
splittable sets arises in this way. Thus this construction produces at
most $2^{|I|}$ alignments without splittable sets. Further alignments
with the same point matches may be obtained by merging mergeable sets
as described in Sect.~\ref{sec:ambiguous}.

The admissible choices can be determined efficiently. An implied match
requires two set-matches of $\mathcal{N}^{\star}$, each of which is
either fixed, i.e., contained in $\mathcal{N}(\mathcal{M})$, or determined
by the orientation of a single isolated match. Representing the
orientation of each isolated match by a Boolean variable, the
admissibility conditions are therefore unary and binary constraints,
i.e., a 2-SAT instance, whose satisfiability can be decided in linear time
\cite{Aspvall:79} and whose solutions can be enumerated by backtracking
over the orientation variables with unit propagation. In the absence of
label constraints, moreover, every choice of orientations for a
\emph{maximal} consistent set $\mathcal{M}$ is admissible: if an implied
match $(g_1,g_2)\notin\mathcal{M}$ arose from the set-matches of one or
two orientations, then $g_1$ and $g_2$ would be unmatched points, and a
direct check of (C1) and (C2) along the lines of
Sect.~\ref{sec:maxcommon} shows that $\mathcal{M}\cup\{(g_1,g_2)\}$ is
consistent, contradicting the maximality of $\mathcal{M}$. Only if the
candidate matches are restricted by labels (Sect.~\ref{sec:constraints})
can orientations be excluded, as in the example of
Fig.~\ref{fig:Egral}(b), in which the point match $\{4,d;2,b\}$ that
would be implied by the alternative orientation is not part of
$\mathcal{M}$. For strictly monotone scoring functions that depend on
point matches only, all admissible orientations of $\mathcal{M}$ have the
same score. Choices among co-optimal alignments can nevertheless
affect subsequent progressive steps.

We summarize the discussion of the pairwise case:
\begin{proposition}\label{prop:correct}
  Let $\mathcal{P}_0\subseteq E_1\times E_2$ be the set of allowed
  candidate matches and let $f$ be a scoring function that is strictly
  monotone on sets of allowed matches. Then:
  \begin{description}
  \item[(i)] A set $\mathcal{M}\subseteq\mathcal{P}_0$ is the set of
    matches of an alignment of $(E_1,\mathscr{V}_1)$ and
    $(E_2,\mathscr{V}_2)$ if and only if $\mathcal{M}$ is consistent,
    closed under implied matches, and admits an admissible choice of
    orientations for its isolated matches.
  \item[(ii)] The search tree of $\mathrm{Match}(\emptyset,\mathcal{P}_0)$,
    Alg.~\ref{alg:BKstyle1}, with compatibility defined as consistency,
    contains every consistent $\mathcal{M}\subseteq\mathcal{P}_0$ as a
    node. A score-optimal alignment is therefore obtained by evaluating
    $f$ at every node that satisfies the conditions of (i) and retaining
    the best one, together with an admissible choice of orientations.
  \item[(iii)] If $\mathcal{P}_0=E_1\times E_2$, every maximal consistent
    $\mathcal{M}$ is closed, every choice of orientations is admissible,
    and hence the optimum is attained at a leaf of the search tree. For
    $f(\mathcal{M})=|\mathcal{M}|$, the optimal alignments correspond to
    the maximum common edge subgraphs of the corresponding graphs.
  \end{description}
\end{proposition}
\begin{proof}
  (i) The set of matches of an alignment is consistent by
  Lemma~\ref{lem:matchsub} and Prop.~\ref{prop:consistent}, closed by
  the first part of the proof of Lemma~\ref{lem:implied}, which does not
  use the absence of isolated matches, and its sets determine an
  admissible orientation. The converse is Lemma~\ref{lem:implied}
  together with the discussion of orientations above. (ii) follows from
  the discussion in Sect.~\ref{sec:maxcommon}: if $\mathcal{M}$ is
  consistent and $p$ is its last element in the fixed candidate order,
  then $\mathcal{M}\setminus\{p\}$ is consistent and, by induction, a node
  of the search tree at which $p$ is still a candidate and compatible,
  so that $\mathcal{M}$ is one of its children. (iii) Closedness of
  maximal consistent sets and admissibility of all orientations were shown
  above. Since $f$ is strictly monotone and every consistent set is
  contained in a maximal one, the optimum is attained at a maximal
  consistent set, i.e., at a leaf. The correspondence with maximum common
  edge subgraphs is Prop.~\ref{prop:MCES}.
\end{proof}

\subsection{Progressive Alignments}\label{sec:progressive}

The decomposability of alignments also makes it possible to partition
alignments recursively along a binary tree. Each internal node of this
``\textit{guide tree}'' then corresponds to a pairwise alignment of its
children.  Alignments of E-graphs, like vertex-based alignments of ordinary
graphs, alignments of trees, and alignments of sequences, therefore can be
constructed \emph{progressively}, i.e., by means of stepwisely computing
pairwise alignments of input E-graphs, an input E-graph and an E-graph
alignment, or two E-graph alignments. Since E-graphs alignments are again
E-graphs, algorithmically there is, in principle, no difference between the
three cases.  As noted above, the construction of pairwise E-graph
alignments does not introduce splittable sets. Intermediate alignments
  may nevertheless contain mergeable sets, and merging them can change the
  pairwise search space at a later step. Fixing an intermediate alignment
  therefore fixes one representative among these alternatives.

Let $(E_1,\mathscr{V}_1)$ and $(E_2,\mathscr{V}_2)$ be E-graph alignments
with disjoint sets of rows $S_1$ and $S_2$, respectively.  If $e_1$ and
$e_1'$ are incident in $E_1$ while $e_2$ and $e_2'$ are not incident in
$E_2$, it is still possible to obtain an alignment with point matches
$(e_1,e_2)$ and $(e_1',e_2')$ provided there is a pair of mergeable sets
$\Upsilon$ and $\Upsilon'$ in $E_2$ such that $e_2\in\Upsilon$ and
$e_2'\in\Upsilon'$.  That is, two non-incident points that are incident
with mergeable sets in one E-graph can be matched with pair of incident
points in the E-graph. 
A search that also allows alternative representatives would therefore
need to consider the following additional cases:
\begin{itemize}
\item[(iv)] $e_1$ and $e_1'$ are incident in $E_1$, $e_2$ and $e_2'$ are
  not incident in $E_2$, and there is a pair of mergeable sets $\Upsilon_2$
  and $\Upsilon_2'$ with $e_2\in\Upsilon_2$ and $e_2'\in\Upsilon_2'$, or
\item[(v)] $e_2$ and $e_2'$ are incident in $E_2$, $e_1$ and $e_1'$ are not
  incident in $E_1$, and there is a pair of mergeable sets $\Upsilon_1$ and
  $\Upsilon_1'$ with $e_1\in\Upsilon_1$ and $e_1'\in\Upsilon_1'$.
\end{itemize}

Point matches do not uniquely determine all admissible merges of
ambiguous sets. The progressive procedure considered here retains the
current intermediate representatives and does not explore alternative
merges. Pairwise optimality is therefore restricted to these fixed
representatives; it does not imply optimality over alternative merges,
earlier co-optimal choices, or input orders.

Progressive alignments in general will not reach the globally optimal
score. This is well known for sequence alignments \cite{Feng:87}.  This
observation immediately carries over to E-graphs, since a string/sequence
can be understood as a path E-graph with each character of the input string
corresponding to a labeled edge in the E-graph.

\begin{algorithm}
  \caption{Procedure $\textrm{PMEGA}(\{G_1,G_2,\dots,G_N\})$}
  \label{alg:BKstyle}
  \begin{algorithmic}
    \STATE compute binary guide tree $T(G_1,G_2,\dots,G_N)$ 
    \FORALL{inner nodes $u$ of $T$ in bottom-up order}
      \STATE $\{v,w\}\leftarrow \text{children}(u)$
      \STATE $G_u\leftarrow \text{OptimalAlignment}(G_v,G_w)$ 
    \ENDFOR 
  \end{algorithmic}
  \label{alg:PMEGA}
\end{algorithm}

The simple progressive alignment procedure $\textrm{PMEGA}()$ is detailed
in Alg.~\ref{alg:PMEGA}. The pairwise alignment of two E-graphs, $G'$ and
$G''$, is computed by $\textit{OptimalAlignment}(G',G'')$ for the two
fixed representatives. The subroutine maximizes the score over
consistent, closed point matchings with jointly admissible,
label-compatible endpoint orientations and constructs the corresponding
alignment E-graph. Any rooted binary tree with the
input E-graphs as leaves can serve as guide tree. In analogy to
progressive sequence alignment \cite{Feng:87}, a natural choice is a
tree obtained by hierarchical clustering, e.g.\ UPGMA, from the pairwise
alignment scores; the simplest choice is the caterpillar tree defined by
the input order, in which the input E-graphs are added one at a time to a
growing alignment. The implementation described in
Appendix~\ref{sec:implementation} uses the latter, i.e.,
Alg.~\ref{alg:multiple_alignment} is the specialization of
Alg.~\ref{alg:PMEGA} to this particular guide tree.

\begin{figure}[h]
	\begin{center}
		\includegraphics[width=0.7\columnwidth]{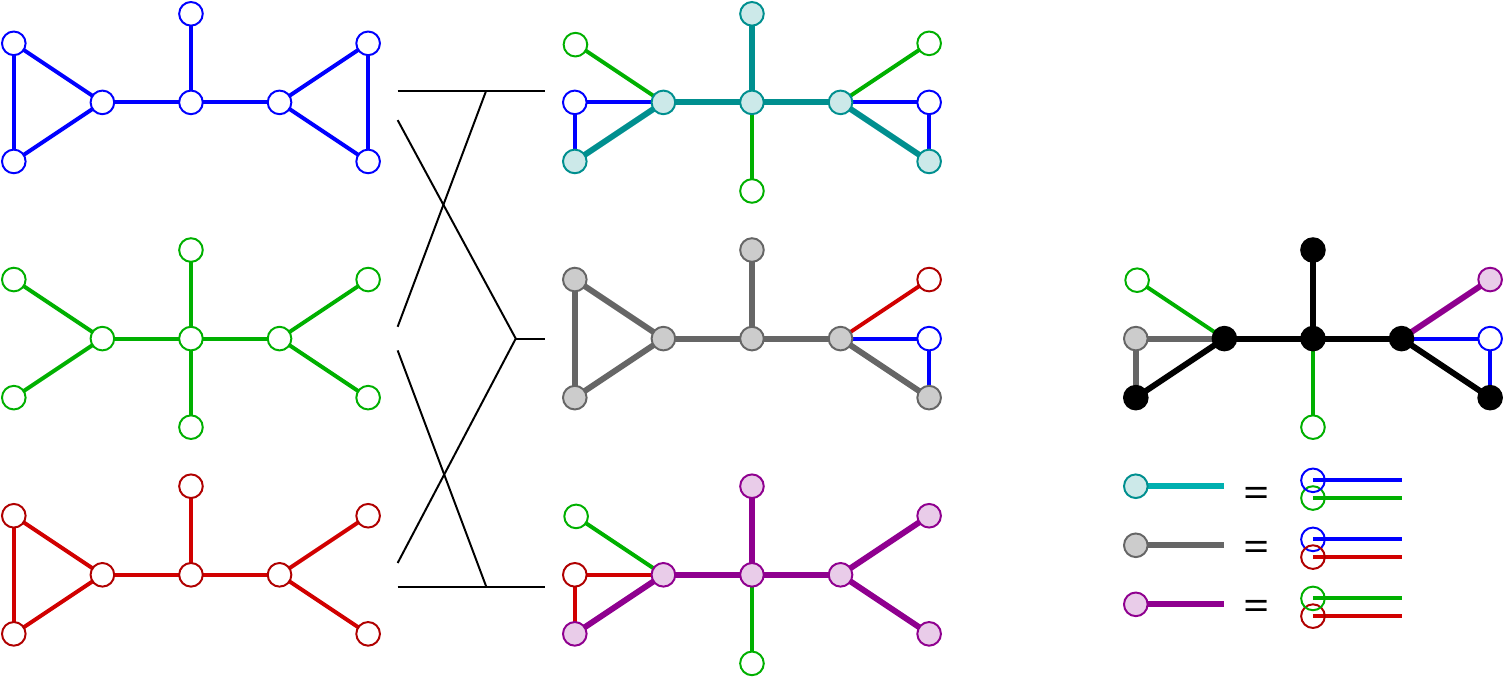}
	\end{center}
	\caption{Multiple (three-way) alignment (to the right) of the three
		graphs shown in different colors in the l.h.s.\ column.  The middle
		column shows the three pairwise alignments that are contained in the
		three-way alignment. Match columns (i.e., edges in the alignment graph)
		are shown in black, all implied vertex matches are shaded. In/del edges
		are shown in the color of the graph that defines the column. In the
		three-way alignment, the content of the alignment columns is also
		color-coded. (black: match all three input graphs, cyan = blue \&
		green; gray = blue \& red; magenta = green \& red). Note that the
		three-way alignment is the pairwise alignment of an input graph and a
		pairwise alignment.}
	\label{fig:PMEGA} 
\end{figure}

A worked example of PMEGA in action, showing how three input E-graphs are progressively merged into a composite alignment, is illustrated in Fig.~\ref{fig:PMEGA}.

Alignments of alignments are naturally scored using a (possibly normalized)
sum-of-pairs score as commonly employed in sequence alignments
\cite{Carillo:88}: For each column, the scores for all pairs of input
graphs $G_i$ and $G_j$ contained in the multiple alignments $G'$ and $G''$
are summed or averaged. More precisely, let $S'$ and $S''$ be the
row sets of $G'$ and $G''$. By Eq.~\eqref{eq:SP}, the contribution of a
point match $(e',e'')$ of a column $e'$ of $G'$ and a column $e''$ of
$G''$, relative to leaving both columns unmatched, is
$\sum_{\alpha,\beta}\big[\sigma(\Lambda_\alpha(e'),\Lambda_\beta(e''))-2\gamma\big]$,
where the sum runs over all $\alpha\in S'$ present in $e'$ and all
$\beta\in S''$ present in $e''$; the terms of Eq.~\eqref{eq:SP} involving
two rows of the same input alignment are the same for all alignments of
$G'$ and $G''$. Thus the sum-of-pairs score induces a scoring function
for the pairwise alignment of $G'$ and $G''$ that is strictly monotone
under the same condition $\sigma(x,y)>2\gamma$ as in the pairwise case,
with the feasible-state search of Prop.~\ref{prop:correct} applying
when hard label constraints are present.

\subsection{Additional Constraints on Points and Sets}\label{sec:constraints}

The definition of compatible points and sets, i.e., of compatible vertices
and edges in the corresponding graphs, can be constrained to user-defined
combinations of labels.  In a chemical context, for example, one might want
to suppress matches between hydrogen and carbon atoms or between bonds of
different order. Constraints on edge matches are easily accommodated in
Algorithm \ref{alg:BKstyle1} by restricting the set of candidate matches
$\mathcal{P}$. 

Let $\mathcal{P}_0\subseteq E_1\times E_2$ denote the initial set
of label-allowed candidate matches, independently of the current search
frontier $\mathcal{P}$. Both forbidden edge matches and forbidden vertex matches may lead to
conflicts for implied edges. These are easily handled, however: It suffices
to check in Algorithm \ref{alg:BKstyle1} that $J\subseteq\mathcal{P}_0$, i.e., that
every implied match is an allowed candidate for a match. For every implied
match $p'\in J$, one checks in addition whether the implied vertex matches,
i.e., the relative orientation of the match is allowed. For example,
suppose it may be allowed in general to match $C$ with $N$, but $C-N$ and
$C=N$ can only be matched so that the two carbons match. Even though an implied $(C-N,C=N)$ match may be in $\mathcal{P}_0$, it cannot be inserted as
$(N-C,C=N)$ match whenever the delimiting NC and CN matches are already
given in $\mathcal{M}$.  Algorithm \ref{alg:BKstyle1} thus must be modified such that
it checks, for all $p'\in J$, whether $\mathcal{M}\cup\{p\}\cup\{p'\}$ is
compatible. Note that the insertion of an implied point match cannot force
additional implied point-matches or set-matches as long as all constraints
concern only points and sets. If $J\not\subseteq\mathcal{P}_0$ or there is $p'\in J$
such that $\text{compatible}(\mathcal{M}\cup\{p\},p')$ returns false, then the
corresponding branch of the search tree must be abandoned since
$\mathcal{M}\cup\{p\}$ is not compatible.
Label constraints therefore change the relationship between alignments
and common edge subgraphs: a maximum cardinality consistent set of label-compatible
matches, i.e., a solution of the labeled MCES problem, is not necessarily
closed under implied matches and then does not extend to an
alignment. The optimal E-graph alignment is the best closed consistent
set of allowed matches that admits an admissible choice of orientations
(Prop.~\ref{prop:correct}), and it may therefore comprise fewer matches
than a labeled MCES. By Prop.~\ref{prop:correct}(ii), the backtracking
procedure nevertheless finds it, provided that the score is evaluated at
every node of the search tree that satisfies these conditions, rather
than at the leaves only.

In the presence of constraints such as the \{C-N, C=N\} example above, as
well as in the case of oriented graph edges, some isolated point matches
may already imply a unique correspondence of their endpoints.  These can
immediately be inserted into the data structures holding the match
information $\mathcal{M}$, i.e., into the set matches $\mathcal{N}$. Note, however,
that technically, there is no set in the E-graph associated with this
match, even though the match of the vertices in the associated graph is uniquely
defined.

Set-dependent score contributions, i.e., scores for (mis)matches of
vertex labels, can be accommodated as follows. Let
$\ell_\alpha:\mathscr{V}_\alpha\to\mathscr{L}_V$ assign labels to the
sets, let $\tau:\mathscr{L}_V\times\mathscr{L}_V\to\mathbb{R}$ score a
(mis)match of set labels, and let $\gamma_V\le 0$ be the score of an
unmatched set. 
For a pairwise alignment without splittable sets, the total score
with point matches $\mathcal{M}$ and set-matches
$\mathcal{N}^{\star}$, including
those fixed by the orientations of isolated matches, is then
\begin{equation}\label{eq:setscore}
	f(\mathcal{M})+\sum_{(\Upsilon_1,\Upsilon_2)\in\mathcal{N}^{\star}}
	\tau\big(\ell_1(\Upsilon_1),\ell_2(\Upsilon_2)\big)
	+\gamma_V\,\big(|\mathscr{V}_1|+|\mathscr{V}_2|-2|\mathcal{N}^{\star}|\big)\,.
\end{equation}
An extension that preserves all existing set-matches increases the
total score if $\sigma(x,y)>2\gamma$ and $\tau(a,b)\ge 2\gamma_V$ for
all allowed labels. In general, however, adding a point match may
reorient previously isolated matches and replace existing set-matches.
These inequalities therefore do not justify restricting the search to
cardinality-maximal point matchings. General set-dependent scoring
requires evaluating every feasible point matching and optimizing its
admissible orientations for Eq.~\eqref{eq:setscore}.

\subsection{Modifications to the VF2-like Algorithms} 

The matching routine in the VF2 scheme is modified in
Alg.~\ref{alg:BKstyle2} by a routine
$\textrm{Complete}(\mathcal{M},\mathcal{N},p)$, Alg.~\ref{alg:complete},
that attempts to augment the set $\mathcal{M}$ of current
point-matches and its associated set $\mathcal{N}$ of set matches by
candidate point-match $p$. Instead of just adding $p$, all implied
point-matches are inserted by
$\textrm{complete}(\mathcal{M},\mathcal{N},p)$. In addition, the set
$\mathcal{N}$ of set matches is updated. The routine returns the
  symbol $\bot$ if $\mathcal{M}\cup\{p\}$ is not consistent or if an
  implied match is not an allowed candidate. Lines 1--9 of
  Alg.~\ref{alg:complete} constitute the consistency test of
  Def.~\ref{Def:consistent}: condition (C1) only needs to be checked for
  the pairs involving $p$, and condition (C2) only for the new set-matches
  $\mathcal{N}'$, which are obtained by applying (N1) and (N2) to the sets
  containing $e_1$ and $e_2$ and to the sets containing the matched points
  incident with them. By the remark at the beginning of
  Sect.~\ref{sec:maxcommon}, the implied matches collected in $J$ do not
  give rise to further set-matches or implied matches, so that a single
  pass suffices. For the fixed-representative progressive variant, these
  checks are applied to the current representatives without additional
  ambiguous-set merges, as specified in Sect.~\ref{sec:progressive}.

\begin{algorithm}
  \caption{Procedure $\textrm{Match}(\mathcal{M},\mathcal{N},\mathcal{P})$}
	\label{alg:BKstyle2}
	\begin{algorithmic}[1]
          
            \IF{$\mathcal{M}$ admits jointly admissible, label-compatible orientations
            }
            \STATE maximize the score over these orientations and retain
            a copy of the best oriented state if it improves the incumbent
            \ENDIF
            \IF{$\textrm{Complete}(\mathcal{M},\mathcal{N},p)=\bot$ for all $p\in\mathcal{P}$}
            \STATE \textbf{return}
            \ENDIF
            \FORALL{$p\in \mathcal{P}$ in a fixed order}
            \STATE $\mathcal{P}\leftarrow \mathcal{P}\setminus \{p\}$
            \STATE $(\mathcal{M}',\mathcal{N}')\leftarrow\textrm{Complete}(\mathcal{M},\mathcal{N},p)$
            \IF{$(\mathcal{M}',\mathcal{N}')\ne\bot$}
            \STATE $\text{Match}\big(\mathcal{M}',\mathcal{N}',\textrm{update}(\mathcal{P},\mathcal{M}')\big)$
            \ENDIF
            \ENDFOR
	\end{algorithmic} 
\end{algorithm}

The routine $\textrm{update}(\mathcal{P},\mathcal{M}')$ updates the
list of candidate matches by removing all potential point-matches that
contain a point in one of the input E-graphs that is contained in one of
the point-matches that have been added to $\mathcal{M}'$.

\begin{algorithm}
  \caption{Procedure $\textrm{Complete}(\mathcal{M},\mathcal{N},p)$ for
      a consistent set $\mathcal{M}$ with $\mathcal{N}=\mathcal{N}(\mathcal{M})$
      and a candidate $p=(e_1,e_2)$}
  \label{alg:complete}
  \begin{algorithmic}[1]
    \FORALL{$q=(e_1',e_2')\in\mathcal{M}$}
      \IF{exactly one of ``$e_1,e_1'$ incident in $E_1$'' and ``$e_2,e_2'$ incident in $E_2$'' holds}
      \STATE \textbf{return} $\bot$ \COMMENT{(C1) violated}
      \ENDIF
      \ENDFOR
      \STATE $\mathcal{N}'\leftarrow\mathcal{N}(\mathcal{M}\cup\{p\})\setminus\mathcal{N}$
      \COMMENT{new set-matches by (N1) and (N2)}
      \IF{$\mathcal{N}\cup\mathcal{N}'$ is not a label-compatible partial bijection}
      \STATE \textbf{return} $\bot$ \COMMENT{(C2) violated}
      \ENDIF
      \STATE $J\leftarrow\{p\}$
      \FORALL{$(\Upsilon_1,\Upsilon_2)\in\mathcal{N}'$ and
        $(\Upsilon_1',\Upsilon_2')\in\mathcal{N}\cup\mathcal{N}'$ with
        $\Upsilon_1\cap\Upsilon_1'=\{f_1\}$ and $\Upsilon_2\cap\Upsilon_2'=\{f_2\}$}
      \STATE $q\leftarrow(f_1,f_2)$ \COMMENT{implied match}
      \IF{$q\notin\mathcal{M}\cup J$}
      \IF{the label combination of $q$ is not allowed}
      \STATE \textbf{return} $\bot$
      \ENDIF
      \STATE $J\leftarrow J\cup\{q\}$
      \ENDIF
      \ENDFOR
      \STATE \textbf{return} $(\mathcal{M}\cup J,\ \mathcal{N}
              \cup\mathcal{N}')$
	\end{algorithmic} 
\end{algorithm}

\subsection{Complexity}\label{sec:complexity}
Since the MCES problem is NP-hard \cite{Garey:79} and, by
Prop.~\ref{prop:MCES}, a special case of pairwise E-graph alignment, a
polynomial-time exact algorithm for the general pairwise problem would
imply P$=$NP. Let $m_i=|E_i|$ and $m=\min(m_1,m_2)$. The number of
partial point injections is
$\sum_{k=0}^{m}\binom{m_1}{k}\binom{m_2}{k}k!$.
Different explicit extensions may yield the same closed matching, so
completion does not preserve the uniqueness of search states.

A consistent matching of $k$ points has at most $k$
isolated matched components, each with at most two endpoint
orientations. Hence the number of distinct oriented states is bounded by
\[
\sum_{k=0}^{m}\binom{m_1}{k}\binom{m_2}{k}k!\,2^k.
\]
This is a bound on the number of distinct states, not on the number of
recursive visits in the reference prototype: different sequences of
extensions may reach the same state. The count concerns point-based
scoring for fixed representatives;
it excludes additional ambiguous-set merges and general soft
vertex-score optimization. A progressive alignment of $N$ inputs
performs $N-1$ pairwise searches, with intermediate point counts bounded
by the total number of input points.

\subsection{Implementation}
We provide a Python reference prototype for progressive E-graph
  alignment, described in Appendix~\ref{sec:implementation}.
The source code and
  showcase data are supplied as Supplementary Software. The project
  repository is \url{https://github.com/AkbarDavoodi/EdgeAlign}.
Joint search over endpoint orientations, combined with memoization and upper bound pruning (Algorithm~\ref{alg:find_best_alignment}), reduces
the median runtime on a synthetic pair of uniformly labelled seven-edge trees from $16.8$~s for the direct recursive prototype to $0.203$~s for the current implementation (approximately 83-fold; Appendix~\ref{app:prototype-scaling}). This instance-specific speedup reflects the combined implementation changes. Runtime still grows rapidly on the benchmark families. We expect further efficiency gains from exploiting the E-graph incidence structure more systematically.

\section{Applications} 

\subsection{Consensus Structures}\label{sec:consensus}

From a multiple alignment $(E,\mathscr{V})$ of E-graphs we can extract a
consensus E-graph using a suitable voting rule. More formally, for each
$e\in E$ we define a vector of labels $\Lambda(e)=(\Lambda_{1}(e),\ldots,
\Lambda_{|S|}(e))$, in which
\begin{equation}
  \Lambda_{\alpha}(e) = \begin{cases}
    \lambda_{\alpha}(\mu_{\alpha}^{-1}(e))
    & \text{if}\quad \mu_{\alpha}^{-1}(e)\ne\emptyset \\
    \GAP & \text{if}\quad \mu_{\alpha}^{-1}(e)=\emptyset
  \end{cases}
\end{equation}
Here, $\lambda_{\alpha}:E_{\alpha}\to \mathscr{L}$ denotes the edge-label
function for the input E-graph $G_{\alpha}$, and $\GAP\notin\mathscr{L}$ is
the gap symbol denoting the fact that no edge of $G_{\alpha}$ takes
part in the alignment column $e$.

A \emph{voting rule} is then simply a function
$\Phi:(\mathscr{L}\cup\{\GAP\})^{|S|}\to \mathscr{L}'\cup\{\GAP\}$ that
assigns a single symbol from an alphabet $\mathscr{L}'$ or a gap symbol to
each edge of the alignment E-graph depending on the labels of the observed
points (graph edges) in each column. Naturally we assume that
$\mathscr{L}\subseteq\mathscr{L}'$, i.e., every original letter may also
appear as consensus label. In applications of chemical graphs, where points
designate chemical bonds, the pertinent labels are e.g.\ the bond types.

In general the ``consensus alphabet'' $\mathscr{L}'$ may contain more
letters than $\mathscr{L}$. For example, the IUPAC notation for nucleic
acids allows not only the four nucleotides $\mathscr{L}=\{A,G,C,T\}$ but
also combinations such as $N=A|G|C|T$, $R=G|A$, and $Y=C|T$ denoting sets
of nucleotides \cite{Schneider:02}. Similarly, in applications to
chemistry, $\mathscr{L}'$ might include e.g., symbols for ``arbitrary
halogen'' or for ``oxygen or sulfur''. For bond types as example, single,
aromatic, and non-aromatic multiple bonds could be added to
$\mathscr{L}'$. Considering the representation of reactions in the form of
RCLASSes in KEGG \cite{Kotera:04,Kanehisa:14}, bonds within and outside
of rings could be distinguished.

Given a voting rule $\Phi$, the \emph{consensus E-graph} is defined as
\begin{equation}\label{eq:Consensus}
	(E_{\Phi},\mathscr{V}_{\Phi}) :=
	(E,\mathscr{V})[\{e \ | \ \Phi(\Lambda(e))\ne\GAP\}]
\end{equation}
i.e., it is obtained from the alignment E-graph as the sub-E-graph induced
by the edges not voted to carry a $\GAP$ label. A simple definition for
$\Phi(\Lambda(e))$ is to retain $e$ if a certain fraction (e.g.\ half) of
the $\lambda_{\alpha}(\mu_{\alpha}^{-1}(e))\ne \GAP$ and to use the most
abundant label among all retained edges. Of course, more complicated rules
could also be used. For instance, one might want to use a label $X$ for
halogens if none of the halogens appears in the overwhelming
number of cases.

\begin{figure}[h]
  \begin{center}
    \includegraphics[width=0.5\columnwidth]{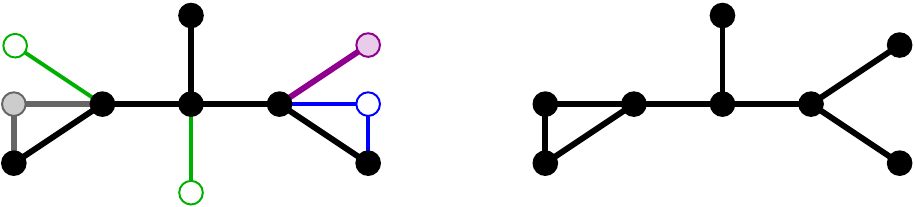}
  \end{center}
  \caption{Majority vote consensus E-graph (r.h.s.) of the multiple
    alignment E-graph of Fig.~\ref{fig:PMEGA} (l.h.s.). An edge is included
    in the consensus if it is present in at least half of the input
    graphs. Thus in this example edges appearing only in one of the three
    input E-graphs are deleted.}
  \label{fig:MSA} 
\end{figure}

Applying a majority-vote rule to the three-way alignment of
Fig.~\ref{fig:PMEGA} yields the consensus E-graph depicted in
Fig.~\ref{fig:MSA}, in which only edges present in at least two of the
three input graphs are retained.

\subsection{A Showcase Example}
To demonstrate the effectiveness of our computational framework in
identifying consensus structures among similar molecular graphs, we
selected five molecules of moderate size and structural similarity as an
illustrative example: \texttt{O=C1CCC(=O)N1},
\texttt{O=C1NC(=O)C(C(=O)O)N1}, \texttt{CC(=O)NC(N)=O},
\texttt{CCC(=O)NC(N)=O}, and \texttt{NC(=O)NC(N)=O}. We applied the
  progressive graph alignment procedure to these molecules to
identify commonly mapped edges and highlight structural consensus through a
simple majority rule.  The molecular graphs were generated from the
  SMILES strings with RDKit and comprise the heavy atoms only, i.e.,
  hydrogens are implicit. Points (bonds) are labeled by the RDKit bond type
  (single or double; none of the five molecules contains aromatic bonds)
  and sets (atoms) by the element symbol. Both labels are used as hard
  constraints, as is customary for labeled MCES problems in cheminformatics
  \cite{Raymond:02,Akbar:27}: a candidate point match is allowed only if
  the bond types agree and the element symbols of the two endpoints agree
  as multisets, and implied matches that violate these constraints prune
  the corresponding branch of the search tree
  (Sect.~\ref{sec:constraints}). The scoring function is the number of
  allowed matches, i.e., $\sigma\equiv 1$ and $\gamma=0$ in
  Eq.~\eqref{eq:pairscore}; for the alignment of the growing composite
  alignment with the next input E-graph the implementation uses the
  lexicographic score $(c(\mathcal{M}),|\mathcal{M}|)$, where
  $c(\mathcal{M})$ is the number of match columns in which all E-graphs
  aligned so far are present (Appendix~\ref{sec:implementation}). The input
  E-graphs were added in the order listed above, i.e., along the
  caterpillar guide tree of Sect.~\ref{sec:progressive}; the pairwise
  choices recorded in Table~\ref{tab:columns} are co-optimal for their
  fixed preceding representatives. The four pairwise scores are $(6,6)$,
  $(5,6)$, $(5,7)$, and $(4,5)$, each optimal for the corresponding fixed
  preceding representative. Alternative co-optimal choices may yield
  different point correspondences. No isolated matched components remain in
  the matchings recorded in the table.  The number of columns in which all
  E-graphs aligned so far are present is $6$, $5$, $5$, and $4$ after the
  four progressive steps, respectively. The consensus was determined with
  the majority rule of Sect.~\ref{sec:consensus}: an alignment column is
  retained if it is present in at least three of the five molecules, and
  columns present in all five molecules are distinguished. The input data,
  implementation, and a machine-readable record of the deterministic
  implementation output are available in the GitHub repository. This
  output agrees with Table~\ref{tab:columns} after the molecule-2 identifier
  translation specified in its caption and a permutation of columns.
\begin{figure}
  \begin{center}
    \includegraphics[width=1\linewidth]{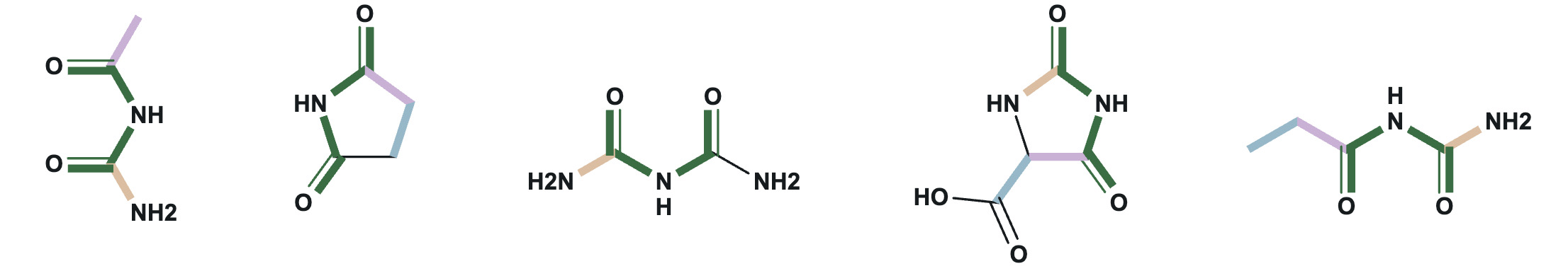}
  \end{center}
  \caption{Visualization of five molecular graphs highlighting
    consensus edges based on alignment results. Fully conserved edges
    across all molecules are shown in green, while edges conserved in at
    least three molecules are marked with light blue, light orange, and
    light purple. From left to right, the panels show molecules 3, 1, 5, 2, and 4 in the
    	numbering of Table~\ref{tab:columns}.}
  \label{fig:molecule_alignment}
\end{figure}

Fig.~\ref{fig:molecule_alignment} illustrates these five molecules with
color-coded edges based on our alignment results. Edges highlighted in
green represent those fully conserved and mapped across all five molecules,
clearly delineating a robust structural core shared
universally. Furthermore, we employed a voting mechanism to identify edges
conserved in a majority of the structures. Edges appearing in at least
three out of the five molecules are highlighted using distinct colors
(light blue, light orange, and light purple), enabling easy visualization
of commonly shared yet not fully universal structural features. This
makes it possible to quantify and visualize structural consistency
among the molecules.

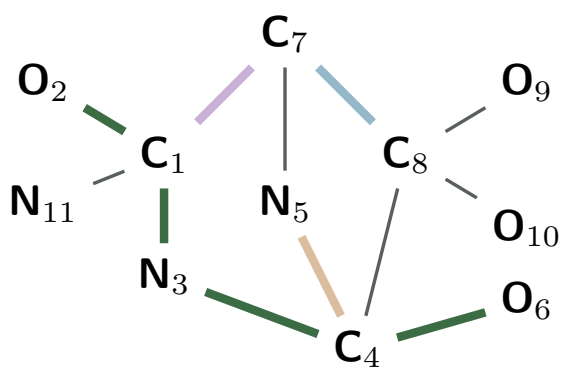
\begin{figure}
	\definecolor{colGreen}{HTML}{3b6c43}
	\definecolor{colPink}{HTML}{c8b0d6}
	\definecolor{colOrange}{HTML}{dbbda0}
	\definecolor{colBlue}{HTML}{96b8cb}
	\definecolor{col}{HTML}{595e5e}
	\begin{center}
		\scalebox{.8}{
			\begin{tikzpicture}[scale=2, every node/.style={draw=none, font=\sffamily\bfseries}]
				
\node (v1) at (0,1) {C$_7$};
\node (v2) at (-.5,.5) {C$_1$};
\node (v3) at (-.5, 0) {N$_3$};
\node (v4) at (0.3,-.3) {C$_4$};
\node (v5) at (.5,.5) {C$_8$};

\node (v7) at (0,.3) {N$_5$};

\node (v8) at (1,.2) {O$_{10}$};
\node (v9) at (1,.8) {O$_9$};

\node (v10) at (1,-.1) {O$_6$};

\node (v6) at (-1,.8) {O$_2$};
\node (v11) at (-1,.3) {N$_{11}$};


\draw[line width=2pt, colPink] (v1) -- (v2);
\draw[line width=2pt, colBlue] (v5) -- (v1);
\draw[line width=2pt, colOrange] (v4) -- (v7);

\draw[thick, col] (v4) -- (v5);

\draw[thick, col] (v1) -- (v7);
\draw[thick, col] (v8) -- (v5)--(v9);

\draw[line width=2pt, colGreen] (v2) -- (v6);
\draw[line width=2pt, colGreen] (v2) -- (v3);
\draw[line width=2pt, colGreen] (v3) -- (v4);
\draw[line width=2pt, colGreen] (v4) -- (v10);

\draw[thick, col] (v2) -- (v11);
				
\end{tikzpicture}
}
\caption{Graph representation of the alignment E-graph computed for the
  five molecules in Fig.~\ref{fig:molecule_alignment}. Consensus structural
  elements are highlighted to show universal (dark green) and prevalent
  alignment patterns; the edges correspond to the alignment columns
    and the vertices to the alignment sets listed in
    Table~\ref{tab:columns}, with vertex labels indicating the common
    element symbol of the aligned atoms. This alignment graph represents
  an abstract superposition and contains information on consensus
  patterns. It is not intended to represent a valid chemical structure.}
\label{fig:full_alignment}
\end{center}
\end{figure}

Fig.~\ref{fig:full_alignment} shows the graph representation of the
alignment E-graph, providing a clear graphical depiction of the structural
correspondences across the molecule set.  It is important to note that this
alignment does not correspond to any actual chemical compound or
molecule. Rather, it is an abstract representation of how the molecular
graphs align according to our method. The figure should be interpreted
as a structural consensus map derived from graph alignment. It is not
intended to represent a valid structure. As a superstructure, it contains
each of the five input molecules in Fig.~\ref{fig:molecule_alignment}
as a subgraph. Table~\ref{tab:columns} lists, for each of the twelve
alignment columns, the bonds of the five molecules that are mapped onto
it and the two alignment sets (vertices of Fig.~\ref{fig:full_alignment})
containing it; the projection of the alignment onto each row thus
recovers the respective molecular graph. The fully conserved core
consists of the four bonds of the imide/urea motif O=C--N--C=O, while the
second nitrogen (column~3) and the C--C bond to the second carbonyl
carbon (column~6) are missing in exactly one molecule each.

\begin{table}
  \caption{Columns of the alignment E-graph of the five molecules of
    Fig.~\ref{fig:molecule_alignment}. Bonds are identified by their
    labels and by their end atoms, numbered in the order of the heavy
    atoms in the SMILES strings. Molecule~2 uses RDKit numbering here:
    repository identifiers
    $(b1,b2,b3,b4,b5,b6,b7,b8,b9,b10)$ correspond to table identifiers
    $(b7,b8,b6,b9,b10,b1,b2,b3,b4,b5)$, respectively; all other
    molecule identifiers agree. ``--'' denotes an in/del. The last two columns give the alignment sets
    (vertices of Fig.~\ref{fig:full_alignment}) containing the column and
    the number of molecules in which the column is present.}
  \label{tab:columns}
  \footnotesize
  \setlength{\tabcolsep}{3.5pt}
  \begin{tabular}{@{}clllllllc@{}}
    \toprule
    No. & bond & molecule 1 & molecule 2 & molecule 3 & molecule 4 & molecule 5 & sets & \# \\
		\midrule
		1 & double & a5 (C5=O6) & b4 (C4=O5)  & c2 (C2=O3) & d3 (C3=O4) & e6 (C5=O7) & C$_1$,O$_2$ & 5\\
		2 & single & a6 (C5--N7) & b3 (N3--C4) & c3 (C2--N4) & d4 (C3--N5) & e4 (N4--C5) & C$_1$,N$_3$ & 5\\
		3 & single & --          & b10 (N10--C2) & c5 (C5--N6) & d6 (C6--N7) & e1 (N1--C2) & C$_4$,N$_5$ & 4\\
		4 & single & a7 (N7--C2) & b2 (C2--N3) & c4 (N4--C5) & d5 (N5--C6) & e3 (C2--N4) & N$_3$,C$_4$ & 5\\
		5 & double & a1 (O1=C2) & b1 (C2=O1)  & c6 (C5=O7) & d7 (C6=O8) & e2 (C2=O3) & C$_4$,O$_6$ & 5\\
		6 & single & a4 (C4--C5) & b5 (C4--C6) & c1 (C1--C2) & d2 (C2--C3) & --          & C$_1$,C$_7$ & 4\\
		7 & single & a3 (C3--C4) & b6 (C7--C6) & --          & d1 (C1--C2) & --          & C$_7$,C$_8$ & 3\\
		8 & single & a2 (C2--C3) & --          & --          & --          & --          & C$_4$,C$_8$ & 1\\
		9 & single & --          & b9 (C6--N10) & --         & --          & --          & N$_5$,C$_7$ & 1\\
		10 & double & --          & b7 (C7=O8)  & --          & --          & --          & C$_8$,O$_9$ & 1\\
		11 & single & --          & b8 (C7--O9) & --          & --          & --          & C$_8$,O$_{10}$ & 1\\
		12 & single & --          & --          & --          & --          & e5 (C5--N6) & C$_1$,N$_{11}$ & 1\\
    \bottomrule
  \end{tabular}
\end{table}

This computational example showcases how our alignment method effectively
captures consensus structures, highlighting both universally conserved
elements and those prevalent enough to be considered structurally
significant across closely related molecular graphs. 

\section{Discussion} 
We have introduced edge-graphs as a set-theoretic structure that enables
the construction of pairwise and multiple alignments. Following the general
theory of abstract alignments \cite{Berkemer:20b}, alignments of
edge-graphs are themselves edge-graphs, from which the input data can be
recovered as row-wise projections. Similar to vertex-based graph
alignments, where alignment matches correspond to common induced subgraphs,
edge-graph alignments are associated with the MCES problem. Traditionally,
MCES is addressed by approximating it as a MCIS problem on the line
graphs. In light of Whitney's result on the almost equivalence of the
isomorphism of graphs and their line graphs \cite{Whitney:32} this yields
very good approximations, particularly when considering the connected
version of MCES, and exact solutions if the $\Delta Y$ exchanges
  corresponding to the exceptional case are excluded explicitly
  \cite{Raymond:02}.  Nevertheless it is of both theoretical and practical
interest to explore exact, direct solutions for this class of problems. The
mathematical framework of edge-graphs enables the construction of a
VF2-like algorithm for computing exact solutions to unlabeled
pairwise MCES problems. With hard label constraints, the alignment
  search instead optimizes over feasible closed matchings, which need not
  coincide with a labeled MCES. Pairwise searches on fixed representatives
  can be combined to provide a heuristic solution to the multiple
alignment problem using the well-known paradigm of progressive
alignments. In terms of its search space, the pairwise procedure of
  Sect.~\ref{sec:maxcommon} is equivalent to a line-graph MCIS search in
  which $\Delta Y$ exchanges are excluded; the advantage of the E-graph
  formalism is therefore conceptual rather than computational. The
  set-matches make the exclusion of $\Delta Y$ exchanges part of the
  consistency test of partial solutions, and, more importantly, the same
  consistency conditions apply to fixed intermediate representatives. In
  the progressive implementation these representatives are retained, so
  alternative ambiguous-set merges and earlier co-optimal choices are not
  exhaustively explored.

The relationship between E-graph alignments and the exact enumeration
  of common subgraphs of multiple labeled graphs \cite{Akbar:27} deserves
  to be made explicit. The match columns of a multiple E-graph alignment in
  which all rows are present form a common edge subgraph of all input
  graphs (Sect.~\ref{sec:egraphalign} together with Prop.~\ref{prop:MCES}),
  and a multiple alignment optimized for the sum-of-pairs objective
  maximizes agreement over row pairs rather than the fully conserved core
  alone.  The present implementation instead uses the lexicographic
  point-based objective specified in Appendix~\ref{sec:implementation} and
  is a progressive heuristic for that objective.  The exact framework of
  \cite{Akbar:27} determines precisely the maximum fully conserved cores,
  including their connected and label-constrained variants, by recursively
  combining maximal cliques of modular products of line graphs; it thus
  provides a natural exact reference for the fully conserved part of a
  progressive E-graph alignment, against which the loss incurred by the
  progressive heuristic can be quantified on instances of moderate
  size. Conversely, the alignment adds to the common core the columns
  shared by only a subset of the inputs, as well as the in/del columns,
  which is what makes the consensus construction of
  Sect.~\ref{sec:consensus} possible. The anchored MCES used in
  \cite{Akbar:27} to infer graph transformation rules from sets of chemical
  reactions suggests a related application of consensus E-graphs, in which
  the conserved reaction center plays the role of the anchor.

To demonstrate the feasibility of this approach, we implemented this
approach in Python. The main purpose of the present contribution
is to introduce the formal framework of edge-graphs and their alignments
and to provide a reference prototype. We expect the explicit incidence
structure of E-graphs to provide a basis for highly scalable implementations;
exploiting this structure for efficient search is a focus of future work.
In particular, it will be interesting to see to what extent
alternatives to a VF2-like enumeration of pairwise common sub-edge-graphs
can achieve superior performance. A natural contender is the McSplit
algorithm \cite{McCreesh:17,Hoffmann:17,Liu:23} for MCIS that efficiently
prunes the search tree; the clique-based enumeration with similarity-based
ordering of the inputs and pruning of the modular product described in
\cite{Akbar:27} is another. Moreover, alternatives to the simple progressive
scheme for multiple alignments are of immediate interest.


\section*{Declarations}
\section*{Funding}
This work was funded by the Novo Nordisk Foundation as part of the project
MATOMIC (Mathematical Modelling for Microbial Community Induced Metabolic
Diseases, 0066551) and from the European Unions Horizon Europe Doctoral
Network programme under the Marie-Sk{\l}odowska-Curie grant agreement No
101072930 (TACsy - Training Alliance for Computational systems chemistry).
PFS acknowledges the financial support by the Federal Ministry of Education
and Research of Germany (BMBF) through DAAD project 57616814 (SECAI, School
of Embedded Composite AI), and jointly with the S{\"a}chsische
Staatsministerium f{\"u}r Wissenschaft, Kultur und Tourismus in the
programme Center of Excellence for AI-research \emph{Center for Scalable
  Data Analytics and Artificial Intelligence Dresden/Leipzig}, project
identification number: SCADS24B.  The funders had no role in study design,
data collection and analysis, decision to publish, or preparation of the
manuscript.
\section*{Competing interests}
The authors declare that they have no competing interests.
\section*{Ethics approval and consent to participate}
Not applicable
\section*{Consent for publication}
Not applicable
\section*{Data availability}
 The showcase SMILES strings, E-graph inputs, alignment implementation,
 	machin -readable deterministic output, and molecular drawing helpers,
 	together with the Appendix~\ref{app:prototype-scaling} benchmark inputs,
 	frozen implementations, protocol, and timing records, are available in
 	the GitHub repository \url{https://github.com/AkbarDavoodi/EdgeAlign}.
\section*{Materials availability}
Not applicable
\section*{Code availability}
The full source code for the implementation
is publicly available at
\url{https://github.com/AkbarDavoodi/EdgeAlign}.
\section*{Author contributions}
All authors have contributed equally to this work.

\begin{appendices}

\section{Implementation and Further Practical Considerations}

\label{sec:implementation}

We represent an E-graph as a pair $(E,\mathscr{V})$, where $E$
denotes the set of edges (serving as the basic elements or ``points'') and
$\mathscr{V}$ is a collection of sets encoding the incidence
structure (i.e., the vertices defined implicitly by the sets of incident
edges). The overall alignment is represented as a triple
$(E^*,\mathscr{V}^*,\mathcal{N})$, where $E^*$ is the composite
aligned edge set, $\mathscr{V}^*$ is a collection of alignment sets
(clusters of matched edges), and $\mathcal{N}$ is a mapping that
records the correspondence between the incidence sets of the input
E-graphs, i.e., the set-matches of
  Sect.~\ref{sec:egraphalign}. Throughout the Appendix, the notation
  follows the main text: $\mathcal{M}$ denotes a set of point matches and
  $\mathcal{N}$ the corresponding set-matches.

\begingroup
\makeatletter
\let\originalfloatmakebox\float@makebox
\def\float@makebox#1{\color@vbox\normalcolor\originalfloatmakebox{#1}\color@endbox}
\makeatother
\begin{algorithm}[t]
	\caption{\emph{MultipleEGraphAlignment}($\{(E_i,\mathscr{V}_i)\}_{i=1}^n$)}
	\begin{algorithmic}[1]
		\REQUIRE A collection of E-graphs $\{(E_i,\mathscr{V}_i)\}_{i=1}^n$
		\ENSURE Composite alignment $(E^*,\mathscr{V}^*,\mathcal{N})$ 
		\STATE $(E^*,\mathscr{V}^*,\mathcal{N}) \gets (E_1,\mathscr{V}_1,\varnothing)$
		\FOR{$i = 2$ \TO $n$}
		\STATE $d \gets i$
		\COMMENT{number of rows of the composite alignment after this step}
		\STATE $(E^*,\mathscr{V}^*,\mathcal{N}) \gets$
		\textsc{FindBestAlignment}$((E^*,\mathscr{V}^*),(E_i,\mathscr{V}_i),d)$
		\ENDFOR
		\RETURN $(E^*,\mathscr{V}^*,\mathcal{N})$
	\end{algorithmic}
	\label{alg:multiple_alignment}
\end{algorithm}
\endgroup

The implementation follows the caterpillar guide tree of
  Alg.~\ref{alg:multiple_alignment}, retaining each intermediate
  representative. The pairwise search implements Algorithm~\ref{alg:find_best_alignment}. The pairwise routine maintains an injective endpoint map
  $B$ and explores both orientations of each candidate point match.  Known
  point and endpoint labels are hard constraints. Endpoints of an isolated
  edge are stored as two distinct records, even when their singleton
  incidence sets coincide; suppressing both records recovers the canonical
  E-graph convention.

  For an endpoint map $B$, \textsc{Close} includes every edge pair whose
  endpoints correspond under $B$ and rejects any forbidden overlap.
  Endpoint injectivity and label compatibility are checked before a trial
  is accepted. Thus orientations are resolved jointly during the search,
  before a state is retained as an optimum.

  The score is $(c(\mathcal{M}),|\mathcal{M}|)$ in lexicographic order,
  where $c(\mathcal{M})$ counts columns present in all rows. Every feasible
  oriented state, including the empty matching, is scored. Repeated states
  are memoized, candidates are traversed in a fixed order, and an upper
  bound prunes states that cannot improve the incumbent. The bound adds the
  smaller number of unused all-row points to $c(\mathcal{M})$ and the
  smaller number of unused points to $|\mathcal{M}|$.

  Columns are stored as nested tuples, with depth $d$ determining gap
  padding. Incidence sets are stored as \texttt{frozenset} objects and
  endpoints are indexed to distinguish identical singleton stars.  The
  saved optimum is copied before all unmatched points, including
  disconnected components, are added as in/del columns. Auxiliary label
  maps and records for unmatched incidence sets are omitted from the
  pseudocode and from the mathematical relation $\mathcal{N}^*$.
	
\begin{algorithm}
  \caption{\emph{FindBestAlignment}($(E_1,\mathscr{V}_1)$,
    $(E_2,\mathscr{V}_2)$, $d$) for fixed representatives}
  \label{alg:find_best_alignment}
  \begin{algorithmic}[1]
    \REQUIRE Two fixed representatives, their hard labels, and output depth $d$
      \ENSURE A feasible alignment $(E^*,\mathscr{V}^*,\mathcal{N}^*)$
      \STATE $\mathrm{bestScore}\gets(-1,-1)$;
      $(\mathcal{M}^*,B^*)\gets(\varnothing,\varnothing)$;
      $\mathrm{visited}\gets\varnothing$
      \STATE \textbf{procedure} \textsc{Visit}$(B)$:
      \STATE \hspace{1em} $\mathcal{M}\gets\textsc{Close}(B)$
      \STATE \hspace{1em} \textbf{if} $\mathcal{M}=\bot$ or
      $B\in\mathrm{visited}$ \textbf{then return}
      \STATE \hspace{1em} add $B$ to $\mathrm{visited}$
      \STATE \hspace{1em} $\mathrm{score}\gets(c(\mathcal{M}),|\mathcal{M}|)$
      \STATE \hspace{1em} \textbf{if} $\mathrm{score}>\mathrm{bestScore}$
      \textbf{then} save a copy of $(\mathcal{M},B)$ and update
      $\mathrm{bestScore}$
      \STATE \hspace{1em} \textbf{if} a valid score upper bound is at most
      $\mathrm{bestScore}$ \textbf{then return}
      \FORALL{unused, label-allowed pairs $(e_1,e_2)$ in fixed order}
      \FORALL{the two endpoint orientations of $(e_1,e_2)$}
      \STATE extend $B$ by the chosen endpoint correspondence to obtain $B'$
      \IF{$B'$ is an injective, endpoint-label-compatible map}
      \STATE \textsc{Visit}$(B')$
      \ENDIF
      \ENDFOR
      \ENDFOR
      \STATE \textbf{end procedure}
      \STATE \textsc{Visit}$(\varnothing)$
      \STATE $(E^*,\mathscr{V}^*,\mathcal{N}^*)\gets$
      glue the copied best oriented state and add every unmatched point as
      a gap-padded column of depth $d$
      \RETURN $(E^*,\mathscr{V}^*,\mathcal{N}^*)$
  \end{algorithmic}
\end{algorithm}
	
Gluing the endpoints paired by $B$ and including all unmatched points
  yields a simple alignment: injectivity preserves the endpoints of each
  input, while \textsc{Close} identifies every overlapping edge
  pair. Projection therefore recovers both fixed input representatives.
  Conversely, consider a feasible alignment of these representatives
  without additional ambiguous-set merges. Its endpoint correspondences can
  be followed in successive search extensions, and closure adds only
  matches already present in that alignment. The state remains reachable
  unless a valid upper bound excludes improvement of the incumbent.  The
  routine in Alg.~\ref{alg:find_best_alignment} therefore optimizes
  the stated point-based score for the
  fixed representatives. This guarantee does not extend to alternative
  ambiguous-set merges, previous progressive choices, or soft vertex-score
  objectives.

\clearpage

\section[Runtime of the reference prototype]{Runtime of the reference prototype}
\label{app:prototype-scaling}

We compare the direct recursive prototype, with the endpoint-orientation search of Algorithm~\ref{alg:find_best_alignment} on the same synthetic tree pairs. The direct prototype resolves isolated-match orientations after the search and has neither memoization nor upper-bound pruning. The comparison therefore evaluates the combined effect of these implementation changes. Each input pair consists of  the path on vertices $0,\ldots,n$ and the tree obtained by replacing $\{n-1,n\}$ with $\{1,n\}$. All edge labels agree; vertex labels are uniform or repeat C/N/O in vertex-index order. We report medians of three runs and extend each series from $n=3$ until its first 60-second timeout.

\begin{figure}[!ht]
\color{black}
\def\scalingcolor{black}
\centering
\begin{minipage}{0.49\textwidth}\centering
\begin{tikzpicture}
\begin{semilogyaxis}[
width=0.98\linewidth,height=6.3cm,
xmin=2.7,xmax=11.5,ymin=0.00005,ymax=1000,xtick={3,5,7,9,11},
ytick={0.0001,0.01,1,60},yticklabels={$10^{-4}$,$10^{-2}$,1,60},
xlabel={Edges per input ($n$)},ylabel={Runtime (s)},
title={Uniform vertex labels},
axis lines=left,axis line style={\scalingcolor},tick style={\scalingcolor},
every tick label/.append style={font=\scriptsize,text=\scalingcolor},
every axis label/.append style={font=\footnotesize,text=\scalingcolor},
title style={font=\footnotesize,text=\scalingcolor,yshift=6pt},clip=false]
\addplot[black!65,thick,dashed,mark=*,mark size=1.2pt,mark options={solid}]
coordinates {(3,0.002283375) (4,0.01510475) (5,0.15314837) (6,1.5574628) (7,16.824819)};
\node[anchor=south east,font=\scriptsize,text=black!65]
at (axis cs:11,100) {Direct: timeout ($n=8$)};
\addplot[\scalingcolor,thick,mark=*,mark size=1.2pt,mark options={solid}]
coordinates {(3,0.00024404199) (4,0.001052917) (5,0.005953208) (6,0.033844666) (7,0.20250054) (8,1.2419722) (9,7.6594367) (10,48.240725)};
\node[anchor=south east,font=\scriptsize,text=\scalingcolor]
at (axis cs:11,350) {Alg. 6: timeout ($n=11$)};
\addplot[\scalingcolor,thin,densely dotted]
coordinates {(2.7,60) (11.35,60)};
\end{semilogyaxis}
\end{tikzpicture}
\end{minipage}\hfill\begin{minipage}{0.49\textwidth}\centering
\begin{tikzpicture}
\begin{semilogyaxis}[
width=0.98\linewidth,height=6.3cm,
xmin=2.7,xmax=15.5,ymin=0.00005,ymax=1000,xtick={3,6,9,12,15},
ytick={0.0001,0.01,1,60},yticklabels={$10^{-4}$,$10^{-2}$,1,60},
xlabel={Edges per input ($n$)},ylabel={Runtime (s)},
title={Repeating C/N/O labels},
axis lines=left,axis line style={\scalingcolor},tick style={\scalingcolor},
every tick label/.append style={font=\scriptsize,text=\scalingcolor},
every axis label/.append style={font=\footnotesize,text=\scalingcolor},
title style={font=\footnotesize,text=\scalingcolor,yshift=6pt},clip=false]
\addplot[black!65,thick,dashed,mark=*,mark size=1.2pt,mark options={solid}]
coordinates {(3,0.000825375) (4,0.00187725) (5,0.013128709) (6,0.097671958) (7,0.29017529) (8,4.1385496) (9,35.593555)};
\node[anchor=south east,font=\scriptsize,text=black!65]
at (axis cs:15,100) {Direct: timeout ($n=10$)};
\addplot[\scalingcolor,thick,mark=*,mark size=1.2pt,mark options={solid}]
coordinates {(3,0.00011966599) (4,0.00017741698) (5,0.00053250001) (6,0.001414791) (7,0.003074292) (8,0.015172208) (9,0.046058958) (10,0.105209) (11,0.549383) (12,1.7573665) (13,4.1216286) (14,23.72597)};
\node[anchor=south east,font=\scriptsize,text=\scalingcolor]
at (axis cs:15,350) {Alg. 6: timeout ($n=15$)};
\addplot[\scalingcolor,thin,densely dotted]
coordinates {(2.7,60) (15.35,60)};
\end{semilogyaxis}
\end{tikzpicture}
\end{minipage}
\par\smallskip
{\footnotesize\begin{tikzpicture}[baseline=-0.5ex]
\draw[black!65,thick,dashed] (0,0)--(0.65,0);\end{tikzpicture}\ Direct recursive search
\qquad\begin{tikzpicture}[baseline=-0.5ex]
\draw[\scalingcolor,thick] (0,0)--(0.65,0);\end{tikzpicture}\ Endpoint-orientation search (Alg.~6)}
\caption{Scaling on identical tree pairs. The dotted line is the
60-second limit; direct labels identify the first timeout in each series.}
\label{fig:prototype-scaling}
\end{figure}
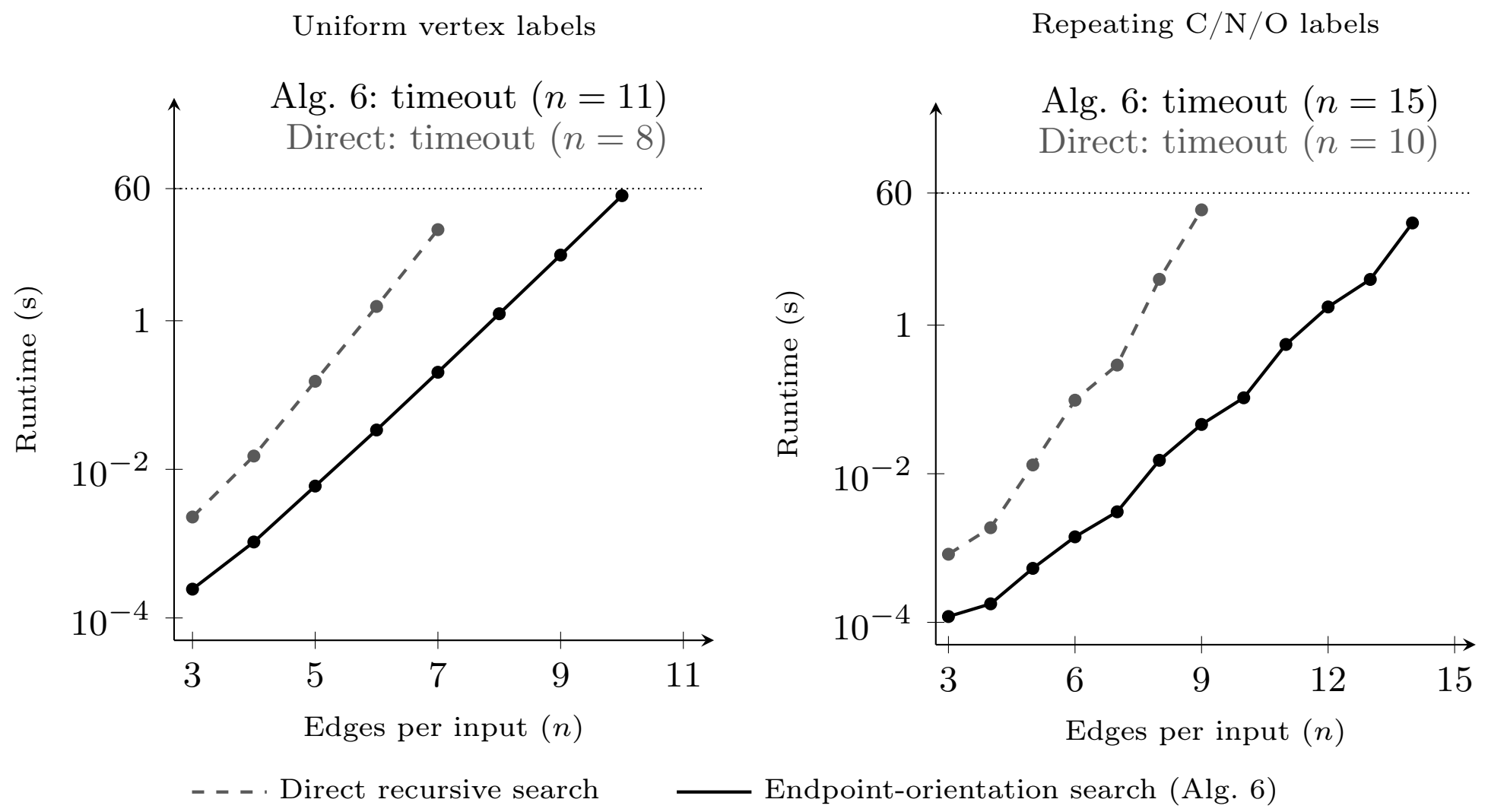

On these families, endpoint-orientation search extends the completed range from 7 to 10 edges per input with uniform labels, and from 9 to 14 with repeating C/N/O labels. Runtime still grows rapidly, leaving further efficiency improvements as a focus of future work.

\end{appendices}


\bibliography{bibliography}

\end{document}